\documentclass[journal]{IEEEtran}
\usepackage{tikz}
\usepackage{amsmath}
\usepackage{cite}
\usepackage{amsmath,amssymb,amsfonts}
\usepackage{algorithmic}
\usepackage{graphicx}
\usepackage{textcomp}
\usepackage{xspace}
\usepackage{xcolor}
\usepackage{multirow}
\usepackage{caption}
\usepackage{subcaption}
\usepackage{float}
\usepackage{pifont}
\usepackage{colortbl}
\usepackage{wasysym}
\usepackage{fontawesome}
\usepackage{booktabs, siunitx}
\usepackage{color, colortbl}
\usepackage{arydshln}
\usepackage{longtable}
\usepackage{supertabular}
\usepackage{arydshln}
\usepackage{url}
\usepackage{tcolorbox}

\usepackage[linesnumbered,vlined]{algorithm2e}

\definecolor{Gray}{gray}{0.9}
\definecolor{LightCyan}{rgb}{0.88,1,1}
\definecolor{Color001}{RGB}{148,0,85}
\definecolor{Color002}{RGB}{0,120,120}
\definecolor{Color003}{RGB}{199,232,172}

\RestyleAlgo{ruled}

\newcommand{\gc}{\cellcolor{Gray}}

\newcommand{\circI}{{\scriptsize\Circle}}
\newcommand{\circII}{{\scriptsize\LEFTcircle}}
\newcommand{\circIII}{{\scriptsize\CIRCLE}}

\usepackage{listings, xcolor}

\definecolor{verylightgray}{rgb}{.97,.97,.97}

\lstdefinelanguage{Solidity}{
	keywords=[1]{anonymous, assembly, assert, balance, break, call, callcode, case, catch, class, constant, continue, constructor, contract, debugger, default, delegatecall, delete, do, else, emit, event, experimental, export, external, false, finally, for, function, gas, if, implements, import, in, indexed, instanceof, interface, internal, is, length, library, log0, log1, log2, log3, log4, memory, modifier, new, payable, pragma, private, protected, public, pure, push, require, return, returns, revert, selfdestruct, send, solidity, storage, struct, suicide, super, switch, then, this, throw, transfer, true, try, typeof, using, value, view, while, with, addmod, ecrecover, keccak256, mulmod, ripemd160, sha256, sha3}, 
	keywordstyle=[1]\color{blue}\bfseries,
	keywords=[2]{address, bool, byte, bytes, bytes1, bytes2, bytes3, bytes4, bytes5, bytes6, bytes7, bytes8, bytes9, bytes10, bytes11, bytes12, bytes13, bytes14, bytes15, bytes16, bytes17, bytes18, bytes19, bytes20, bytes21, bytes22, bytes23, bytes24, bytes25, bytes26, bytes27, bytes28, bytes29, bytes30, bytes31, bytes32, enum, int, int8, int16, int24, int32, int40, int48, int56, int64, int72, int80, int88, int96, int104, int112, int120, int128, int136, int144, int152, int160, int168, int176, int184, int192, int200, int208, int216, int224, int232, int240, int248, int256, mapping, string, uint, uint8, uint16, uint24, uint32, uint40, uint48, uint56, uint64, uint72, uint80, uint88, uint96, uint104, uint112, uint120, uint128, uint136, uint144, uint152, uint160, uint168, uint176, uint184, uint192, uint200, uint208, uint216, uint224, uint232, uint240, uint248, uint256, var, void, ether, finney, szabo, wei, days, hours, minutes, seconds, weeks, years},	
	keywordstyle=[2]\color{teal}\bfseries,
	keywords=[3]{block, blockhash, coinbase, difficulty, gaslimit, number, timestamp, msg, data, gas, sender, sig, value, now, tx, gasprice, origin},	
	keywordstyle=[3]\color{violet}\bfseries,
	identifierstyle=\color{black},
	sensitive=false,
	comment=[l]{//},
	morecomment=[s]{/*}{*/},
	commentstyle=\color{black}\ttfamily,
	stringstyle=\color{red}\ttfamily,
	morestring=[b]',
	morestring=[b]"
}

\newcommand{\sd}{\textit{$\sigma$-deterministic}\xspace}
\newcommand{\snd}{\textit{$\sigma$-nondeterministic}\xspace}

\definecolor{nickgreen}{HTML}{03A60D}
\definecolor{dgreen}{HTML}{03A60D}

\newcommand{\rev}[1]{{\color{black} #1}}

\ifCLASSINFOpdf
\else
\fi
\begin{document}
%
\title{TxT: Real-time Transaction Encapsulation for Ethereum Smart Contracts}
%
%
%

\author{Nikolay~Ivanov,~\IEEEmembership{Graduate Student Member,~IEEE,}
         Qiben~Yan,~\IEEEmembership{Senior~Member,~IEEE}, and Anurag Kompalli%
}

\maketitle

\begin{abstract}
Ethereum is a permissionless blockchain ecosystem that supports execution of smart contracts, the key enablers of decentralized finance (DeFi) and non-fungible tokens (NFT). However, the expressiveness of Ethereum smart contracts is a double-edged sword: while it enables blockchain programmability, it also introduces security vulnerabilities, i.e., the exploitable discrepancies between expected and actual behaviors of the contract code. To address these discrepancies and increase the vulnerability coverage, we propose a new smart contract security testing approach called transaction encapsulation. The core idea lies in the local execution of transactions on a fully-synchronized yet isolated Ethereum node, which creates a preview of outcomes of transaction sequences on the current state of blockchain. This approach poses a critical technical challenge --- the well-known time-of-check/time-of-use (TOCTOU) problem, i.e., \rev{the assurance that the final transactions will exhibit the same execution paths as the encapsulated test transactions.}

In this work, we determine the exact conditions for guaranteed execution path replicability of the tested transactions. To demonstrate the transaction encapsulation, we implement a transaction testing tool, TxT, which reveals the actual outcomes (either benign or malicious) of Ethereum transactions. To ensure the correctness of testing, TxT deterministically verifies whether a given sequence of transactions ensues an identical execution path on the current state of blockchain. We analyze over 1.3 billion Ethereum transactions and determine that 96.5\% of them can be verified by TxT. We further show that TxT successfully reveals the suspicious behaviors associated with 31 out of 37 vulnerabilities (83.8\% coverage) in the smart contract weakness classification (SWC) registry. In comparison, the vulnerability coverage of all the existing defense approaches combined only reaches 40.5\%.
\end{abstract}

\begin{IEEEkeywords}
smart contracts, security, testing
\end{IEEEkeywords}

%
\IEEEpeerreviewmaketitle

\section{Introduction}\label{sec:introduction}
Ethereum smart contracts have been used for a wide variety of decentralized applications, such as decentralized finance (DeFi), non-fungible tokens (NFT), alternative currencies (based on ERC-20 tokens), and data attestation. However, numerous vulnerabilities and attacks on Ethereum smart contracts have been hampering their widespread adoption~\cite{expl1, expl2}.


Following the common vulnerabilities and exposures (CVE) database, the smart contract weakness classification and test cases (SWC) registry~\cite{swcregistry} identifies 37 classes of known smart contract vulnerabilities (as of January 2022). To counter the security threats, different types of defense tools have been developed, including syntactic analyzers~\cite{mavridou2019verisolid,schneidewind2020ethor}, security scanners based on symbolic execution~\cite{kalra2018zeus,luu2016making}, fuzzing tools~\cite{ferreira2021confuzzius,jiang2018contractfuzzer}, transaction analyzers~\cite{chen2020soda,rodler2018sereum}, security libraries~\cite{openzeppelin-contracts,ivanov2021rectifying}, formal defense methods~\cite{cecchetti12compositional,park2018formal}, and various hybrid analysis approaches~\cite{zhou2020ever,ferreira2020aegis}. In this work, we scrutinize 106 existing smart contract security defense solutions, and find that each of them only addresses very few classes of known vulnerabilities. We further discover that certain vulnerability types have never been effectively addressed by any of the proposed defenses.

\rev{
Generally, all the existing smart contract defense methods have two design choices: 1) \emph{heuristic versus deterministic}; and 2) \emph{detection versus verification} (see Table~\ref{tab:determinism}). Heuristic approaches use the best-effort judgement applied to all cases (e.g., Confuzzius~\cite{ferreira2021confuzzius}, sFuzz~\cite{nguyen2020sfuzz}, Harvey~\cite{wustholz2020harvey}), while deterministic designs guarantee the correctness at the expense of rejecting a small number of cases (such as KEVM~\cite{hildenbrandt2018kevm}, SeRIF~\cite{cecchetti12compositional}, and eThor~\cite{schneidewind2020ethor}). Detection tools identify known vulnerabilities (e.g., Oyente~\cite{luu2016making}, Securify~\cite{tsankov2018securify}), while verification tools aim at confirming various safety properties (examples are VerX~\cite{permenev2020verx} and ZEUS~\cite{kalra2018zeus}).
The only known deterministic verification approach is formal verification which proves the correctness of smart contracts by developing formal specifications for an automated prover\cite{chen2020survey}. Unfortunately, these specifications cover only particular cases (e.g., reentrancy~\cite{cecchetti12compositional}). 
Consequently, these formal verification approaches, despite guaranteed correctness, have very limited vulnerability coverage. To increase vulnerability coverage, we propose a new approach for real-time \emph{deterministic verification} of Ethereum transactions.
}

In this work, for the first time, we propose the deterministic verification of Ethereum transactions using a fully-synchronized instrumented Ethereum Virtual Machine (EVM). Our verification system relies on the user confirmation of a test transaction, as smart contract users generally have reasonable expectations of the transaction outcomes. For example, if the users purchase some tokens, they would expect a balance increase of the respective token in the wallet. Unlike traditional defense methods, our approach could cover a large scope of suspicious transactions, thereby revealing the behaviors associated with a majority of known and unknown vulnerabilities.

\begin{table}
\setlength{\tabcolsep}{4pt}
    \centering
    \caption{Different design choices of smart contract defense.
    }
    \label{tab:determinism}
    \begin{tabular}{l||cc:cc}
        
        \arrayrulecolor{red}\toprule
        
        \multicolumn{1}{c}{\multirow{12}{*}{\textbf{Property}}} &
        \multicolumn{4}{c}{\textit{Design choices}} \\
        \cmidrule(lr){2-5}
        

        \textbf{}
        & \multirow[b]{6}{*}{ \rotatebox[origin=l]{90}{\hspace{-7pt} Heuristic~~~~~~~~~}} 
        & \multirow[b]{6}{*}{ \rotatebox[origin=l]{90}{\hspace{-7pt} \textbf{Deterministic}$^\dagger$}}
        & \multirow[b]{6}{*}{ \rotatebox[origin=l]{90}{\hspace{-7pt} Detection~~~~~~~}}
        & \multirow[b]{6}{*}{ \rotatebox[origin=l]{90}{\hspace{-7pt} \textbf{Verification}$^\dagger$~~~}} \\
        
        & & & & \\
        & & & & \\
        & & & & \\
        & & & & \\
        & & & & \\
        
        \arrayrulecolor{red}\midrule
        

        Reject option & \ding{56} & \ding{52} & --- & --- \\
        \hline
        
        Guaranteed correctness & \ding{56} & \ding{52} & --- & --- \\
        \hline
        
        Confirm safety & --- & --- & \ding{56} & \ding{52} \\
        \hline
        
        Identify vulnerabilities & --- & --- & \ding{52} & \ding{56} \\
        
        \arrayrulecolor{red}\bottomrule
        \multicolumn{5}{l}{$^\dagger$ choices made in this work (TxT)}
    \end{tabular}
\end{table}

\noindent\textbf{TxT: Transaction Testing.} 
To make it possible to preview the result of one or several transactions, we develop a smart contract testing framework called \emph{transaction encapsulation}, which uses a fully-synchronized Ethereum node to execute transactions, while preventing the propagation of these transactions across the network. Transaction encapsulation classifies the transactions into two categories: \sd (with guaranteed test result), and \snd (with non-guaranteed test result). To demonstrate the transaction incapsulation, we implement a distributed real-time transaction tester called TxT, which successfully reveals the unexpected outcomes associated with the majority of known smart contract vulnerabilities --- significantly outperforming all existing defense methods. Our evaluation shows that TxT exhibits a low rate of \snd transactions. To further reduce the rate of \snd transactions, we enhance TxT functionality to enable explicit detection of specific vulnerabilities in 75\% of \snd transactions.

To interact with the transaction framework, the  user first connects their crypto wallet to a TxT network and submits a transaction (or a sequence thereof) to the smart contract. 
Then, the user observes in the wallet or dApp interface (if used) the exact outcome of the transaction(s), called \emph{a posteriori state}, manifested in cryptocurrency balances, token balances, error messages, etc. If the result of the test execution matches the expectations,
the user switches their wallet back to the Ethereum Mainnet and submits the transaction as usual.
While the user is testing and submitting transactions, TxT is continuously checking in the background if the condition for the replicability of the test transaction execution path still satisfies.
Without the necessity to install new software or learn contract programming, TxT allows everyday users to identify unexpected outcomes of transaction sequences associated with the majority of known vulnerabilities, and it achieves a high vulnerability coverage which more than doubles
the coverage of all the state-of-the-art defense tools combined.


In summary, we deliver the following contributions:

\begin{itemize}
    \item We propose a new deterministic approach for smart contract verification, \emph{transaction encapsulation}, and design a distributed real-time dynamic transaction tester, TxT, to verify the security of transactions at runtime.
    
    \item To address the time-of-check/time-of-use (TOCTOU) problem, we formally determine the exact set of conditions for the execution path replicability of a test transaction and implement TxT using a fully-synchronized Ethereum node to perform the transaction encapsulation. 
    
    \item We reproduce 37 known smart contract vulnerabilities and confirm that TxT can intercept 83.8\% of them, compared to only 40.5\% by all the existing methods combined. We further evaluate 1.3 billion Ethereum transactions and confirm that 96.5\% of them are suitable for security evaluation by TxT.
\end{itemize}
\section{Background}\label{sec:background}
\noindent\textbf{Ethereum, dApps, and Wallets.} 
Ethereum is a decentralized blockchain ecosystem that supports the  execution of smart contracts. Ethereum popularized the notion of decentralized application (dApp) --- a full-stack software product with a web or mobile interface as a frontend and smart contract as a backend. In order for a dApp to interface with a smart contract and the Ethereum network at large, it must use a wallet as an intermediary. 
The wallets securely store private key(s) for signing and submitting transactions on the user's behalf.

\noindent\textbf{Smart Contracts and Transactions.}
The Ethereum Virtual Machine (EVM) is a part of Ethereum that executes smart contracts. As each transaction is executed by the EVM, the state of the blockchain changes to reflect the executed transaction. However, if a given transaction is invalid, the EVM reverts the blockchain to the state preceding this transaction. Essentially, an Ethereum transaction is a state changing instruction 
signed by the sender using their private keys.



\noindent\textbf{London Hard Fork and EIP-1559.}
There have been instances where Ethereum transactions were included in the blocks paying very little or no gas at all. As of block 12,965,000, a hard fork implementing several new Ethereum features was activated on the network. Dubbed ``London'', this hard fork changed how fees are collected by the Ethereum network. Ethereum Improvement Proposal 1559 (EIP-1559), enforced in the London fork, changes the fee model in a way that it practically prevents zero-priced transactions.

\section{Motivating Example}\label{sec:motivation}

\begin{figure}[t]
    \centering
\begin{lstlisting}[language=Solidity]
contract Foo {
  function deposit() public payable {}
  function withdraw() public {
    address admin = 
    0xEc125A03C6F9E75BEB1A420e94d655B2f1352584;
    payable(admin).transfer(1000000000 wei);
    payable(msg.sender).transfer(address(this).balance);
  }
}
\end{lstlisting}
    \caption{A smart contract that fails only on Mainnet.}
    \label{fig:motivation}
\end{figure}

\begin{figure}[t]
    \centering
\begin{lstlisting}[language=Solidity]
contract Bar { 
  constructor() public { }
}
\end{lstlisting}
    \caption{A non-payable smart contract deployed on Mainnet at \texttt{0xEc125A03C6F9E75BEB1A420e94d655B2f1352584}. The same address on Ropsten testnet is an 
externally owned account (EOA).}
    \label{fig:case21}
\end{figure}

\rev{
Smart contracts do not operate in isolation; instead, they share with other smart contracts a dynamic blockchain network environment. Moreover, the same blockchain platform can be represented by several public blockchain networks, which sometimes affect the execution of the same smart contract. Consider smart contract \texttt{Foo} in Fig.~\ref{fig:motivation}, which transfers funds to smart contract \texttt{Bar} (Fig.~\ref{fig:case21}). \texttt{Bar} is deployed on Mainnet, but not on Ropsten testnet. Moreover, \texttt{Bar} does not have any payable functions, and therefore it cannot accept incoming Ether. As a result, the transfer in line 6 (Fig.~\ref{fig:motivation}) will fail, reverting the entire transaction --- but only on Mainnet, not on Ropsten. Even if the states of all the variables of contract \texttt{Foo} on Ropsten are identical to their counterparts on Mainnet, the behavior of the \texttt{withdraw()} function will be different. This example demonstrates that the state of blockchain (denoted $\sigma$) is an important factor that determines the outcome of smart contract execution.

Next, we run a set of experiments to determine whether the existing smart contract defense can reveal the failed transfer issue. We confirm that Securify~\cite{tsankov2018securify}, Oyente~\cite{luu2016making}, Mythril~\cite{mueller2018smashing}, Vandal~\cite{brent2018vandal}, and Manticore~\cite{mossberg2019manticore} all fail to detect the issue, although some of them produce unrelated warnings. This example shows that some vulnerabilities might not be detected by the existing defense methods. Moreover, the security evaluation on a testnet does not offer a sufficient reassurance of contract safety. To address these issues, we propose a 
new defense approach for smart contracts based on transaction testing.
Our approach tests a transaction (or a series of transactions) on an isolated fully-synchronized node, and then checks in real time whether the test transaction can replicate exactly the same execution path on Mainnet. Unfortunately, most existing smart contract threat mitigation solutions do not take the state of the current environment into account. The solution proposed in this work tests the current state of smart contracts in the blockchain, thereby providing a more accurate representation of contract behaviors. 
}

\section{Preliminaries}\label{sec:overview}

In this section, we introduce the transaction encapsulation approach, and then give an overview of TxT tester, followed by formal conventions, assumptions, and threat model.

\subsection{System Overview}

\noindent \textbf{Transaction Encapsulation.}
In this work, we propose a new \emph{transaction encapsulation} framework which offers a preview of the result of a transaction against the current state of Mainnet, but without mining the transaction across the network. The transaction encapsulation executes one or a series of transactions on an instrumented node fully-synchronized with the Mainnet network. Unlike testnet simulations and symbolic executions, the transaction encapsulation enables the execution of the transaction on the current state of Mainnet. The transaction encapsulation is designed not only to execute the transaction but also to deterministically reason whether the transaction can be replicated on Mainnet with completely identical execution path.

\begin{figure}
    \centering
    \includegraphics[width=\linewidth]{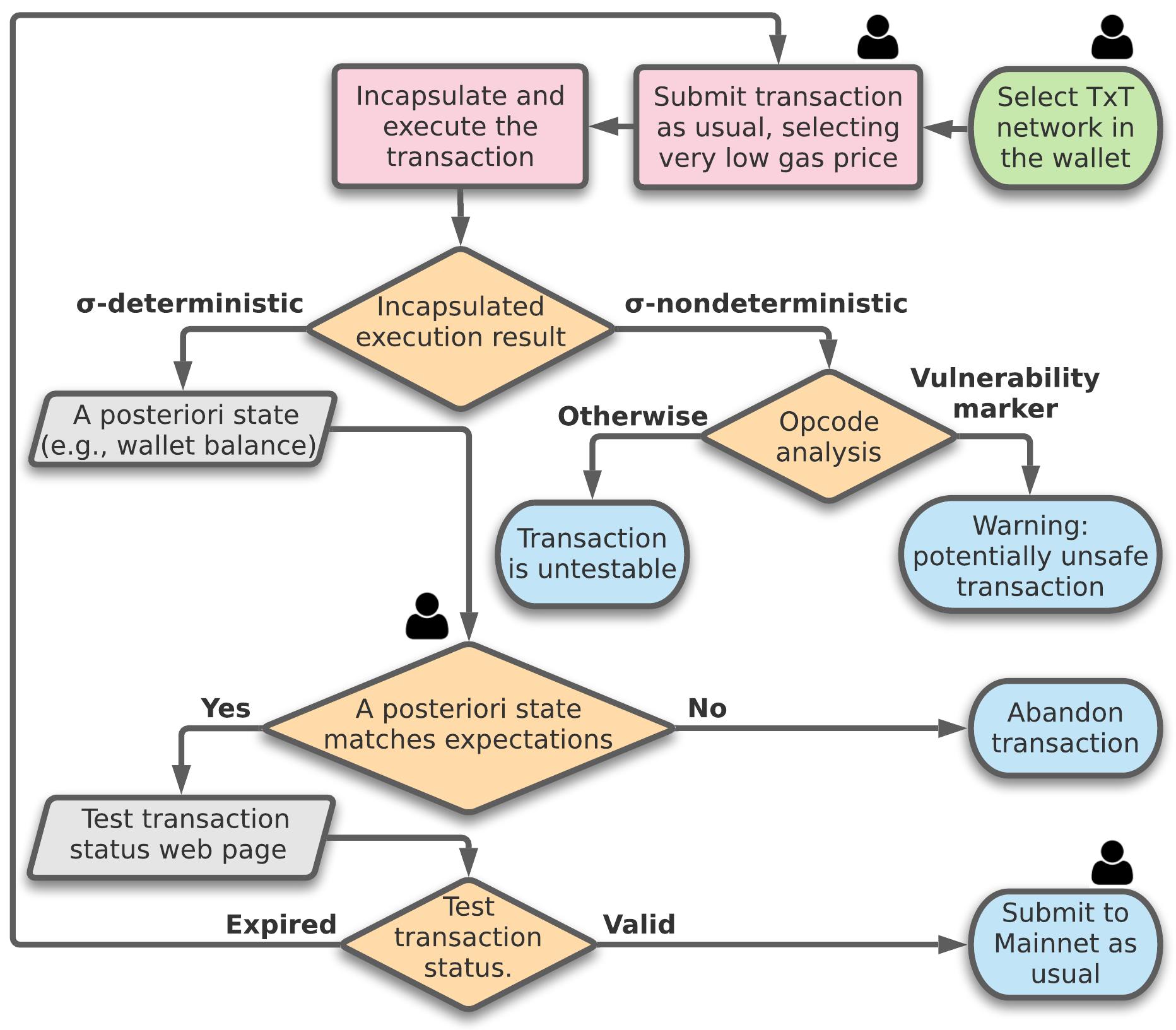}
    \caption{Flow chart of transaction testing. \faUser~--- requires manual user interaction.}
    \label{fig:flow}
\end{figure}

\noindent \textbf{Overview of Transaction Testing Workflow.}
Fig.~\ref{fig:flow} shows the workflow of the TxT's transaction testing. 
To test a transaction with TxT, the user first 
switches the Ethereum network in their wallet and specifies a custom transaction gas price. Then, the user submits a sequence of transactions 
using their favorite wallet and dApp (if applies) --- no other special-purpose software is needed. When the transaction sequence is executed, the \emph{a posteriori state} will be observable in the wallet and/or in the dApp, as if the transaction was executed by the Mainnet. Next, the user observes the status of the tested transaction (e.g., on a web page) to determine if the transaction is testable and reproducible at any given moment.

In some rare cases, TxT will not be able to guarantee the result of the transaction, in which case the transaction will be labelled as \snd. Most \snd transactions contain binary opcodes that are \emph{potentially} associated with some known vulnerabilities --- in this case, TxT issues a warning about such a vulnerability. Otherwise, when the transaction is classified as \snd and there is no vulnerability marker present among the binary opcodes, then the transaction is deemed \emph{untestable}. 

On the other hand, if the transaction is labeled \sd, it means that it is testable \emph{and} guarantees correct test result. In this case, the user observes the result of the transaction (e.g., balances in the wallet) to determine if the result of the transaction matches the expectation. If the result is unexpected, obviously, the transaction should be abandoned by the user. If the result matches the expectation, the user needs to verify whether or not the transaction has \emph{expired}, i.e., whether there are other incoming transactions that change the state of the contract(s) during the transaction testing.
In some rare cases, TxT could determine that the test transaction has expired by the time the user is ready to resubmit it to the Mainnet. Even in such a situation, the user  could retest the transaction. 
Conversely, if a TxT status shows that the transaction is still valid, the user submits the transaction to the Mainnet knowing that the outcome will be identical to the one observed during the corresponding transaction test.

\subsection{Notation}

Previous studies demonstrate that reproducing a smart contract vulnerability often requires a sequence of two or more transactions~\cite{so2021smartest,mossberg2019manticore,nikolic2018finding,krupp2018teether,frank2020ethbmc}. In this work, we use the notation similar to the one in~\cite{so2021smartest} to denote the sequence of $N$ transactions as $T^*$:
\[
T^* = (T_1, \cdots, T_N),\; N \ge 1.
\]
Furthermore, without the loss of generality, we use a simplified\footnote{We simplify the definition by removing fields irrelevant to this study, such as $(v,r,s)$ components of the transaction signature.} notation of transaction adapted from~\cite{wood2014ethereum, web3js-gettransaction}: 
$$T_i = \{T_{n,i}, T_{p,i}, T_{g,i}, T_{o,i}, T_{t,i}, T_{v,i}, T_{f,i}, T_{a,i}, T_{b,i}, T_{h,i}, T_{c,i}\},$$
where $T_{n,i}$ is the transaction nonce, $T_{p,i}$ is gas price, $T_{g,i}$ is gas offer, $T_{o,i}$ is the transaction sender address, $T_{t,i}$ is transaction recipient (destination address), $T_{v,i}$ is the transaction value (the amount of Wei sent along with the transaction), $T_{f,i}$ is the invoked function of the smart contract,
$T_{a,i}$ is the set of arguments with which $T_{f,i}$ is invoked, $T_{b,i}$ is the block the transaction is mined into, $T_{h,i}$ is the transaction hash, and $T_{c,i}$ is the sequence of EVM opcodes in the execution stack of $T_i$, which recursively includes the opcode sequences of all the inter-contract calls (ICCs) executed by the transaction. We assume that $T_i$ is properly signed.



\subsection{Assumptions}

\noindent \textbf{A Posteriori State Assessment.}
Unlike traditional defense methods, TxT does not detect vulnerable or malicious code patterns; instead, TxT reveals \emph{suspicious behavior} associated with these vulnerabilities. Specifically, we make a reasonable assumption that the user can assess whether the outcome of a series of transactions is satisfactory or not.  TxT will then give the user an accurate preview of what will happen if the given transaction sequence is executed, and the user can use the interface of the wallet and/or the dApp to assess the a posteriori state in the form of Ether balances, token balances, dApp interface elements, transaction error messages, etc.

\noindent \textbf{Transaction Sequences.}
We assume that all transactions in the sequence represent a single complete logical workflow, such that the user can unambiguously assess its success or failure. For example, a typical token exchange workflow can be logically represented as the following sequence:
\ding{182} sell token $A$ for stablecoin\footnote{A token with market price pegged to a fiat currency (e.g., USD).} $S$; \ding{183} buy token $B$ using stablecoin $S$. In this example, the user expects to observe a specific amount of B tokens in their wallet. Also, we assume that all transactions in the sequence are distinct and sent from the same account to the same contract, i.e., 
\begin{equation}\label{eq:eq-sequence}
\forall T_i, T_j \in T^*, i \neq j : T_{o,i} = T_{o,j} \wedge T_{t,i} = T_{t,j}, 
\end{equation}
where  $T_i$, $T_j$ are two transactions in the same sequence.  We assume that the transactions in the sequence are chronologically ordered. Since Ethereum uses incremental per-account nonces by design~\cite{wood2014ethereum}, a testable transaction sequence must have nonces appearing in a strictly ascending order, i.e.,
\[
\forall T_i, T_j \in T^* : j = i + 1 \implies T_{n,j} = T_{n,i} + 1. 
\]
Finally, we define the requirement for Ethereum state transitions within the testable transaction sequence:
\begin{equation}\label{eq:eq02}
\begin{array}{l}
\forall T_i, T_j \in T^*: j = i + 1 \wedge T_i \mapsto T_j \implies \nexists T_k : \\
T_k \notin T^* \wedge T_{o,k} = T_{o,i} \wedge T_{n,k} \in [T_{n,1},T_{n,N}],
\end{array}
\end{equation}
where  $T_i \mapsto T_j$ denotes an EVM state transition when transaction $T_j$ is executed after $T_i$ within the sequence, and $\nexists T_k$ indicates the non-existence of any transaction $T_k$ that satisfies the following criteria. 

\noindent \textbf{On-Chain Transactions.}
We assume that all the transactions tested by TxT are traditional \emph{on-chain transactions}, i.e., the transactions propagated, pooled, and mined by unmodified Ethereum nodes, such as \emph{Go-Ethereum}. 
The Decentralized Finance (DeFi) ecosystem, which has gained significant traction in the recent years, is particularly sensitive to transaction ordering manipulation via a widespread opportunistic exploration of Miner/Maximum Extractable Value (MEV)~\cite{mev}. This  creates a pretext for transaction ordering attacks, such as sandwich front-running attack~\cite{ferreira2021frontrunner}. To alleviate the negative consequences (e.g., gas fee inflation and increased network overhead) of MEV transactions, the Flashbots project delivers a patch (MEV-Geth~\cite{mev-geth}) for the Go-Ethereum node that allows DeFi participants to submit transactions directly to the patched nodes, which essentially  creates an off-chain overlay network for transaction propagation.
In this work, we consider orthodox Ethereum transactions, and leave the MEV-related transactions for future work.



\subsection{Threat Model}\label{sec:design-threatmodel}
\rev{
In this work, we assume that Ethereum is secure and correct on the blockchain and consensus layers, and the honest nodes correctly implement the protocol. The threat rests on the smart contract layer, coming either from an attacker or from a non-adversarial bug. The attacker (if present) may either be the one who introduces a security vulnerability in the smart contract, or they may be the one who exploits a pre-existing program bug. The attacker aims at earning financial gains or causing disruptions to the dApps. In all the cases, the attacking vector is a stand-alone Ethereum node or a Ethereum API (such as Infura or Pocket Network).
}


\section{TxT: Transaction Testing Framework}\label{sec:system-design}
In this section, we describe the challenges and details of the TxT design, and illustrate the transaction testing procedure.

\subsection{Design Challenges}

Ethereum is a dynamic ecosystem where anyone in the world can deploy smart contracts or submit transactions that compete for being included into constantly appended blocks. This compositional nature of Ethereum creates a number of practical challenges described below.

\noindent \textbf{Challenge \#1: TOCTOU Problem.}
The time-of-check/time-of-use (TOCTOU) problem is manifested in TxT as the combination of the \emph{transaction expiration problem} and the \emph{execution path guarantee}.
Our analysis of Ethereum confirms the intuitive proposition that the execution path of a transaction does not necessarily repeat that of an identical previously-submitted test transaction. Every test transaction may sooner or later experience an ``expiration'' (i.e., the outcome of the test transaction does not match that of the real transaction), after which it no longer demonstrates a valid outcome of an identical transaction. In this work, we determine the exact set of conditions affecting the expiration of a test transaction, and we further design \emph{TxSEA (Transaction State Expiration Analyzer)} algorithm, 
which could deterministically reason whether a test transaction has expired or not (see Section~\ref{sec:txsea} for more details).
Our analysis of EVM execution reveals that Ethereum smart contracts sometimes include data sources unrelated to transaction-based state transition. For example, the Solidity property \texttt{block.difficulty}, represented by the \texttt{DIFFICULTY} EVM opcode, is determined by mining instead of previous transactions. We call the presence of such data sources \textit{$\sigma$-nondeterminism}. If a transaction exhibits \textit{$\sigma$-nondeterminism} in its execution stack, the transaction is \snd. In this work, we 
determine the exact conditions for \emph{$\sigma$-nondeterminism}, and we design TxT in a way that it unambiguously detects \snd transactions. Moreover, TxT could 
scrutinize \snd transactions to provide a warning regarding specific vulnerabilities associated with the \snd instructions in the contract.

\noindent \textbf{\rev{Challenge \#2: Execution Without Propagation.}} Transaction encapsulation requires that the test transaction should only be executed on the instrumented TxT node, while being ignored by all other nodes within the blockchain network. We show that the straw-man solutions, such as network packet filtering or  propagation suppression of the transaction,  
disrupt the synchronization and lead to a stall of the node. To overcome this challenge, we propose \emph{transaction underpricing} --- a gas price manipulation scheme, which effectively avoids the execution of transaction by the blockchain network at large, without creating conditions in which the TxT node cannot re-synchronize with the Mainnet after the test.


\noindent \textbf{\rev{Challenge \#3: Transaction Sequences.}} As demonstrated by previous studies~\cite{so2021smartest,mossberg2019manticore,nikolic2018finding,krupp2018teether,frank2020ethbmc}, many vulnerabilities require executing a series of transactions for reproduction. To address this challenge, we design TxT to retain the state of a soft fork for a set period of time after each test transaction,  in order to enable the execution of a sequence of transactions with an arbitrary length. We  enhance the TxSEA algorithm to determine the expiration of the entire sequence of transactions.

\subsection{Transaction Expiration}

Determining a transaction expiration event is essential for the success of the proposed TxT tool; otherwise, TxT cannot guarantee that the final real transaction(s) will produce the same result as the test transaction(s).
Here, we formally define the expiration conditions starting from transaction expiration.
\noindent\textbf{Definition 1: }\textit{A transaction $T_i$ is expired at block $B$ if:}
\begin{equation}
\begin{array}{l}
\exists T_j: T_{t,i} = T_{t,j} \wedge T_{o,j} \neq T_{o,i} \wedge T_{b,j} > T_{b,i}
\wedge T_{b,j} \le B. \\
\end{array}
\end{equation}
Essentially, the transaction expiration stipulates the presence of at least one transaction $T_j$ submitted to the same smart contract as $T_i$ ($T_{t,i} = T_{t,j}$) from a different account than $T_i$ ($T_{o,j} \neq T_{o,i}$) at any block time after $T_i$ ($T_{b,j} > T_{b,i}$) but before or at block $B$ ($T_{b,j} \le B$). The following definition asserts that for each block, the sets of expired and unexpired transactions are disjoint and form a partition.

\noindent\textbf{Definition 2: }\textit{A transaction $T_i$ is unexpired at block $B$ if and only if it is not expired at block $B$.}

Following the definitions of transaction expiration, we define the expiration of a transaction sequence as follows.

\noindent\textbf{Definition 3: }\textit{A sequence $T^*$ is expired at block $B$ if:}
\begin{equation}
\begin{array}{l}
\exists T_i \in T^* \; \exists T_j \notin T^*, T_{o,j} \neq T_{o,i} : \\
T_{b,i} < T_{b,j} \le B \wedge T_{t,i} = T_{t,j}. \\    
\end{array}
\end{equation}
Finally, we formally define the condition for an unexpired sequence of transactions.

\noindent\textbf{Definition 4: }\textit{A sequence $T^*$ is unexpired at block $B$ if:}
\begin{equation}
\begin{array}{l}
\forall T_i \in T^* \nexists T_j : \\
T_{o,j} \neq T_{o,i} \wedge T_{b,j} > T_{b,i} \wedge T_{b,j} \le B \wedge T_{t,i} = T_{t,j}. \\    
\end{array}
\end{equation}
Intuitively, a transaction expiration event is characterized by the presence of another transaction calling a function of the same smart contract \emph{after} the test transaction. We assess the probability of such an event in Section~\ref{sec:eval-expiration}.

\subsection{Sources of $\sigma$-nondeterminism}
In order to determine \emph{all} the sources of \emph{$\sigma$-nondeterminism} on the Ethereum platform, we conduct an exhaustive manual analysis of the current 145 EVM opcodes. In the end, we identify the following set of opcodes incurring \emph{$\sigma$-nondeterminism}:
\[
\begin{array}{l}
\mathcal{T} = \{\text{\texttt{BLOCKHASH}},\; \text{\texttt{NUMBER}},\; \text{\texttt{COINBASE}}, \; \text{\texttt{GASLIMIT}}, \\
\text{\texttt{DIFFICULTY}}, \; \text{\texttt{TIMESTAMP}}, \; \text{\texttt{GASPRICE}},\; \text{\texttt{BALANCE}} \}.      
\end{array}
\]
Next, we elaborate on how these opcodes make the associated transaction \snd.

\noindent \textbf{Block Hash.}
The \texttt{BLOCKHASH} opcode retrieves the block hash for a specified block number. Its presence in the execution stack of a transaction is a sign that this transaction is \snd. For example, if $B$ is the most recently mined block, the \texttt{BLOCKHASH} opcode will return \texttt{0x0} for $B+1$ (i.e., the next block). 
However, one hour after that, the same code will return a non-zero hash. Note that the \texttt{BLOCKHASH} opcode constitutes a signature of the ``Weak Sources of Randomness from Chain Attributes'' (SWC-120) vulnerability.

\noindent \textbf{Block Number.}
The \texttt{NUMBER} opcode retrieves the current block number. This variable constantly increments, rendering any transaction that has this opcode in its execution stack to be \snd. Also, this opcode is a marker for the ``Block Values as a Time Proxy'' (SWC-116) vulnerability.

\noindent \textbf{Block Beneficiary Address.}
The block beneficiary address is the address specified by the winning miner for receiving the reward. The \texttt{COINBASE} opcode retrieves the current block's beneficiary address. Since this value may be different between blocks, any transaction that uses this opcode in its execution stack is \snd. Furthermore, this opcode is also a signature of the SWC-120 vulnerability.

\noindent \textbf{Block Gas Limit.}
Each Ethereum block has a limit on the cumulative gas consumption by all its transactions. The \texttt{GASLIMIT} opcode returns the gas limit value. This value may vary from block to block, and therefore the presence of the \texttt{GASLIMIT} opcode within the execution stack of a transaction renders this transaction \snd. Additionally, this opcode constitutes a signature of the SWC-120 vulnerability.

\noindent \textbf{Block Difficulty.}
Each block has its own mining difficulty, which is calculated from the difficulty of the previous block and the timestamp set by the miner, and therefore its specific value is volatile. The \texttt{DIFFICULTY} opcode allows to retrieve the current block's difficulty. The variability of block difficulty is a clear sign that the transaction with the \texttt{DIFFICULTY} opcode in its execution stack is \snd. This opcode is yet another signature of SWC-120.

\noindent \textbf{Block Timestamp.}
The block timestamp is a value put in the block by the miner, and it may not necessarily represent the exact time the block was mined. A contract can retrieve the block timestamp value using the \texttt{TIMESTAMP} opcode.
Intuitively, the value of block timestamp is not expected to stay the same. Therefore, the presence of the \texttt{TIMESTAMP} opcode in the execution stack of a transaction is not only indicative of the SWC-116 vulnerability potential, but it is also an indicator that the transaction is \snd.

\noindent \textbf{Third-party Account Balance.}
The \texttt{BALANCE} opcode retrieves the balance of an account. If an account is not in the set $\{T_{o,i}, T_{t,i}\}$, we call it a third-party account. In this work, we analytically determine that a third-party account balance incurs $\sigma$-nondeterminism in smart contracts. If some account's balance is updated by a transaction submitted to an account other than $T_{t,i}$, it does not render $T_i$ expired; however if this transaction contains a \texttt{BALANCE} opcode in its execution stack, the transaction is marked as \snd.

\noindent \textbf{Transaction Gas Price.}
The transaction gas price can be obtained via the \texttt{GASPRICE} opcode.
Since TxT uses transaction underpricing, 
the value retrieved by the \texttt{GASPRICE} opcode will differ between the test transaction and the final one. Therefore, the presence of this opcode in the execution stack of a transaction implies that this transaction is \snd. This opcode is another signature of the SWC-120 vulnerability.



Finally, by combining the above observations, we can establish the following definitions, starting with the definition of a \sd transaction.

\noindent\textbf{Definition 5: }\textit{A transaction $T_i$ is \sd if and only if $T_{c,i} \cap \mathcal{T} = \varnothing$}.

Since \sd and \snd transactions form a partition, the following definition ensues.

\noindent\textbf{Definition 6: }\textit{A transaction $T_i$ is \snd if and only if it is not \sd.}

Similarly, we can further expand the definitions to include testing sequences.

\noindent\textbf{Definition 7: }\textit{A transaction sequence $T^*$ is \sd if and only if all transaction in $T^*$ are \sd.}


\noindent\textbf{Definition 8:
}\textit{A transaction sequence $T^*$ is \snd if at least one transaction in $T^*$ is \snd.}

\subsection{TxSEA Algorithm}\label{sec:txsea}
Through transaction testing, TxT allows the user to peek into the \emph{a posteriori} state of a transaction. Unfortunately, \emph{a posteriori} state is transient and can expire at any moment.
Due to the other interfering transactions, the execution path of the final transaction might not match that of the testing transaction. To address this issue, we develop the TxSEA algorithm for confirming the identical execution path when the test transaction is submitted to the current block.

\begin{algorithm}[t]
 \KwData{The transaction expiration map $\mathcal{E}$: ContractAddress $\mapsto$ LastTxBlock}
 \SetKwProg{Fn}{Function}{ begin}{end}
 \SetKwProg{Proc}{Procedure}{ begin}{end}
 \Proc{CacheTransaction($T_j$)}{
  \KwResult{Cache the transaction currently processed by EVM and append it to the permanent storage}
  \KwIn{$T_j$ --- currently executed transaction}
  $\mathcal{E}[T_{t,j}] \gets T_{b,j}$\;
 }
 \Fn{ExpirationTest($T_i$)}{
 \KwResult{Test transaction expiration status}
 \KwIn{$T_i$ --- tested transaction}
 \KwOut{\{Expired, Unexpired\}}
  \uIf{$T_{t,i} \notin \mathcal{E}.\text{Keys}$}{
    \Return Unexpired\;
  }
  \uElseIf{$\mathcal{E}[T_{t,i}] \ge T_{b,i}$}{
    \Return Expired\;
  }
  \Else{
    \Return Unexpired\;
  }
 }
 \caption{Dynamic TxSEA with Caching}\label{alg:txsea-dynamic}
\end{algorithm}


Algorithm~\ref{alg:txsea-dynamic} shows an efficient implementation of TxSEA using caching and dynamic programming.
This algorithm introduces a constant-time procedure $CacheTransaction$, which is embedded into the instrumented Ethereum node and invoked for each executed transaction.
This procedure uses the map $\mathcal{E}$ to store the block number of the last transaction for each smart contract. 

The transaction data gathered from the node is stored in an outside storage (e.g., a database), and this data is used by the $ExpirationTest()$ function to determine if the transaction has expired. This function uses the transaction expiration map to search for a transaction that might have been recorded after $T_i$. The condition $\mathcal{E}[T_{t,i}] \notin \mathcal{E}.Keys$ checks whether the smart contract $T_{t,i}$ has any recorded transactions; if not, the transaction is obviously unexpired. Otherwise, we check if the block associated with the last recorded transaction was mined simultaneously or after $T_{b,i}$ (i.e., $\mathcal{E}[T_{t,i}] \ge T_{b,i}$), which indicates expiration. Finally, if the last transaction is recorded for the contract, but it happened before $T_i$, the transaction is unexpired. \rev{Our experiments show that this algorithm only experiences a negligible latency (see Section~\ref{sec:delays}). The requirement for the additional storage does not need an experimental evaluation because it will always occupy a fixed 52 bytes of storage per transaction. As of November 2022, the size of the TxSEA cache is slightly over 86 gigabytes, which is a small fraction of the size of the full node that requires hundreds of gigabytes.}

\subsection{How Does TxT Guarantee the Transaction Execution Path?}
In the previous section, we demonstrate the cases in which TxT cannot guarantee that the execution path of the final Mainnet transaction 
remains the same as that of the test transaction. 
Here, we confirm that, with all the uncertain cases eliminated, the identical path execution can be guaranteed. So far, we have been using a loose notion of transaction execution, which does not take into account the state of blockchain the transaction applies to. Following the Ethereum state transition model from~\cite{wood2014ethereum}, we can further define the formally precise definition of \emph{state-conditional execution path} as follows.

\noindent\textbf{Definition 9: }\textit{A state-conditional execution path, denoted $T_i|\sigma_i$, is the state transition $\sigma_i \rightarrow \sigma'_i$, such that $\sigma'_i = \Upsilon(\sigma_i, T_i$), where $\Upsilon$  is the deterministic state transition function in EVM.}

\noindent\textbf{Definition 10: }\textit{A state of contract $T_{t,i}$, denoted $\sigma_{t,i}$ is a subset of state values in $\sigma_i$ (i.e., $\sigma_{t,i} \in \sigma_i$) that encompass only storage and balances associated with all contracts in the call stack of transaction $T_i$.}

\noindent\textbf{Definition 11: }\textit{A contract-state-conditional execution path with respect to contract $T_{t,i}$, denoted $T_i|\sigma_i|T_{t,i}$, is the state transition $\sigma_{t,i} \rightarrow \sigma'_{t,i}$, such that $\sigma'_{t,i} = \Upsilon(\sigma_{t,i}, T_i)$.}


Now that we have formal definitions of state-conditional execution path, contract state, and contract-state-conditional execution path, consider the following theorem, which formalizes the exact condition of replicability of a transaction execution path.

\noindent\textbf{Theorem 1:} \textit{Given two transactions $T_i$ and $T_j$, if $T_{n,i} = T_{n,j}, T_{g,i} = T_{g,j}, T_{o,i} = T_{o,j}, T_{t,i} = T_{t,j}, T_{v,i} = T_{v,j}, T_{f,i} = T_{f,j}, T_{a,i} = T_{a,j}, T_{c,i} = T_{c,j}, T_{c,i} \notin \mathcal{T}$, and $T_i$ is not expired at block $T_{b,j}$, then $T_i|\sigma_i|T_{t,i} = T_j|\sigma_j|T_{t,j}$, i.e., $T_j$ exhibits an identical execution path as $T_i$ within the call stack of $T_{t,i}$ and conditional to states $\sigma_j$ and $\sigma_i$, respectively, for all $j > i$.}

\noindent\textbf{Proof:}
By definition,  $T_i|\sigma_i|T_{t,i}$ = $T_j|\sigma_j|T_{t,j} \implies \sigma'_{t,i} = \sigma'_{t,j} \implies \Upsilon(\sigma_{t,i}, T_i) = \Upsilon(\sigma_{t,j}, T_j)$.
Since $\Upsilon$ is deterministic, $\sigma'_{t,i}$ depends solely upon $\sigma_{t,i}$ and $T_i$, while $\sigma'_{t,j}$ depends solely upon $\sigma_{t,j}$ and $T_j$.

As per Ethereum and EVM specifications~\cite{wood2014ethereum,evm-opcodes,etherspecs}, the execution of a transaction calling a function of a smart contract is determined only by the following four components: 1) the code of the smart contract, as well as the code of the other contracts invoked within the call stack of the transaction; 2) the storage of the target smart contract, as well as the storage of all contracts within the call stack of the current transaction; 3) balances of smart contracts and EOAs; 4) block-related values. Next, we prove that none of these components could prevent $T_j$, applied to $\sigma_{t,j}$, from executing the exactly same path as $T_i$, when applied to $\sigma_{t,i}$, while satisfying the Theorem's constraints.

Since $T_{c,i} = T_{c,j}$, the code of all contracts in the call stack is identical, and therefore this component is incapable of creating a discrepancy between $T_i|\sigma_i|T_{t,i}$ and $T_j|\sigma_j|T_{t,j}$. As per Definitions 1 and 2, the pre-condition that $T_i$ is not expired at block $T_{b,j}$ implies that $T_{t,i}$ has no incoming transactions to $T_{t,i}$ between the timestamps of blocks $T_{b,i}$ and $T_{b,j}$, i.e.:
\begin{equation}\label{eq:nostor}
\begin{array}{l}
\nexists T_k : T_{o,j} \neq T_{o,i} \implies \\
T_{b,k} > T_{b,i} \wedge T_{b,k} \le T_{b,j} \wedge T_{t,i} = T_{t,k}. \\
\end{array}
\end{equation}
Since $T_{c,i} = T_{c,j}$ and the contract storage can only be altered through an incoming transaction
, Eq.~(\ref{eq:nostor}) effectively eliminates contract storage discrepancy between states $\sigma_{t,i}$ and $\sigma_{t,j}$. Therefore, a contract storage could not create a discrepancy between $T_i|\sigma_i|T_{t,i}$ and $T_j|\sigma_j|T_{t,j}$. 

Similarly, altering a contract's balance is only possible through transactions, mining, or self-destruction. Specifically, the balance increase requires a transaction calling a payable function with a non-zero value. The balance decrease requires a transfer of Ether performed by the smart contract code. The mining reward involves updating of the \emph{coinbase} parameter of the block, which makes it a block-related parameter as discussed later. The self-destruction is only possible by executing the \texttt{SELFDESTRUCT} opcode in the smart contract code initiated via a transaction. Therefore, Eq.~(\ref{eq:nostor}) also excludes any balance transfer. Moreover, since $T_{c,i} \notin \mathcal{T}$, the balance checks for other accounts are also excluded. Therefore, balances also could not create a discrepancy between $T_i|\sigma_i|T_{t,i}$ and $T_j|\sigma_j|T_{t,j}$. Finally, all block-related values are included in $\mathcal{T}$. As established earlier, $T_{c,i} = T_{c,j}$, and thus $T_{c,i} \notin \mathcal{T} \implies T_{c,j} \notin \mathcal{T}$. Therefore, block values cannot create a discrepancy between $T_i|\sigma_i|T_{t,i}$ and $T_j|\sigma_j|T_{t,j}$ under the set constraints.

In summary, we see that the code of the transaction call stack, the storage of the target smart contract and all contracts within the call stack of the current transaction, balances of smart contracts and EOAs, and block-related values are unable to create a discrepancy between $T_i|\sigma_i|T_{t,i}$ and $T_j|\sigma_j|T_{t,j}$. Therefore, $T_i|\sigma_i|T_{t,i} \equiv T_j|\sigma_j|T_{t,j}$.
$\blacksquare$

The set of constraints in Theorem 1 \emph{is the sufficient condition for guaranteed replicability of a test transaction}, which is used by TxT and TxSEA. Specifically, we prove that \sd unexpired transactions guarantee the replicability of a testing transaction execution path.




\subsection{Temporal Separation of Transactions}
Some smart contracts force transaction separation by a time gap. For example, an investment scheme might require a delayed withdrawal of dividends.
In this work, we analytically determine that it is impossible to enforce time separation without incurring $\sigma$-nondeterminism or transaction sequence expiration, which we can summarize in the following theorem.

\noindent\textbf{Theorem 2:} \textit{Inter-transaction time separation stipulation in a sequence $T^*$ implies that $T^*$ is \snd or it is bound to expire before block $T_{b,N}$.}

\noindent\textbf{Proof:} The time separation stipulation means that it is impossible to complete the transaction sequence without awaiting a certain event or condition between a pair of subsequent transactions. Without the loss of rigor, we assume that the minimum inter-transaction separation time quantum is equal to one block\footnote{On average, it takes 10 to 20 seconds in Ethereum to mine a new block.}. This reduction allows us to define the transaction time separation stipulation as follows:
\[
\begin{array}{l}
\exists T_i,T_j \in T^* : T_{b,i} = \alpha \wedge T_{b,j} = \beta \wedge T_{n,j} = T_{n,i} + 1 \\
\implies \beta > \alpha.
\end{array}
\]
This condition is indicative of either of the following three circumstances regarding $T_i$ and $T_j$: 1) The cumulative gas consumption of $T_i$ and $T_j$ exceeds the block's gas limit; 2) There is at least one other transaction $T_k$ expected before the block $\alpha$; 3) The state of blockchain $\sigma$ must meet a certain condition before $\alpha$.
Indeed, outside of these three conditions, there is no other circumstance preventing $T_i$ and $T_j$ with different nonces to share a block. The first condition is automatically prevented by Ethereum by mining one of the transactions in one of the following blocks, but this adaptive behavior is not stipulated because the block size is variable and may or may not exceed the cumulative gas consumption of $T_i$ and $T_j$. The second case precisely matches \emph{Definition 3}, and therefore incurs the expiration of sequence $T^*$. The third case satisfies \emph{Definition 6} (and subsequently \emph{Definition 5}) --- which means that in this case $T_i$ is \snd, and by \emph{Definition 8} it means that $T^*$ is \snd. Therefore, the inter-transaction time separation implies either \emph{$\sigma$-nondeterminism} or expiration of $T^*$ $\blacksquare$


\noindent\textbf{Corollary of Theorem 2:} \emph{If a transaction sequence requires a time separation between transactions, this sequence is untestable, potentially vulnerable, or it will expire before the execution of its last transaction. Therefore, transaction sequences that require time separation between transactions cannot be tested by TxT}.

    

\subsection{Transaction Execution on an Instrumented Node}
Here, we outline some straw-man approaches that might be considered as alternative design choices for TxT. However, all these approaches suffer from some limitations as illustrated below. 

\noindent \textbf{Gossip Delivery.}
TxT requires the user to switch to TxT network for transaction testing. 
It would be reasonable to consider delivering the test transaction through the normal Ethereum node based on the assumption that any transaction, even the one deemed for failure, must arrive at every node in the network, so that all the nodes could make their own independent rejection judgement. However, our experiments show that this assumption is not always correct. Our extensive experiments with transaction underpricing show that the nodes often refuse to forward transactions that do not pass certain ``smoke tests'' (e.g., minimum gas price), so we cannot rely on the Ethereum gossip protocol for delivering the test transaction to the TxT node.

\noindent \textbf{Network-layer Propagation Inhibition.}
We assume that the user has a subscription with a TxT provider. This allows the provider to compare the \texttt{from} field of the transaction message with the user database to filter outgoing network packets containing test transactions.
However, our experiments show that the attempts to tamper with Ethereum network traffic cause some unpredictable behavior, such as node stalling and various synchronization errors. Even if we could overcome these errors by reverse-engineering the software (Go Ethereum, in our case), the reliance on eccentricities of a specific implementation of Ethereum node is not only extremely complicated, 
but  it could also involve some unforeseeable errors. Thus, we choose to prevent the transaction propagation through the less intrusive method of transaction underpricing. Moreover, since the wallets ask users to select the transaction gas price anyway, the requirement to specify a low gas price does not create a noticeable inconvenience.

\noindent \textbf{Submit Final Transaction via TxT.}
If a TxT test confirms the safety of a transaction, the user is required to reconnect to the Mainnet network for submitting the final transaction. This step raises a question: would it be easier to submit the final transaction through TxT instead, as the TxT node is essentially a Mainnet node? Unfortunately, this approach might be less convenient for the user than the one proposed in our design. Submitting the final transaction through TxT would require pruning and re-synchronizing the TxT node to remove the test transaction from it, which takes some time; since tested transactions, as we know, are prone to expiration, any unreasonable delay should be eliminated.

\noindent \textbf{Our TxT Design with Transaction Underpricing.}
TxT requires the isolated execution of transactions on a fully-synchronized node. We run multiple experiments to determine that the ``forced'' solutions, such as gossip firewalling (suppression of transaction propagation), incur unrecoverable node stalls.
Moreover, any extensive modification of a TxT node creates sustainability issues: the same modifications have to be applied to future releases of the node, resulting in an increased maintenance overhead. 
To overcome this challenge, we propose \emph{transaction underpricing} --- a gas price manipulation scheme, which effectively avoids the execution of transaction by the blockchain network at large, without creating conditions in which the TxT node cannot re-synchronize with the Mainnet after the test. To override the rejection of transaction, we enable
\texttt{-{}-miner.gasprice 1} CLI option in Go-Ethereum which effectively overrides the underpriced transaction checks. 
Since the London Fork, Ethereum enforces the EIP-1559 proposal~\cite{buterin2019eip}, which effectively prevents mining transactions with very low gas price, making the concern about accidental mining of underpriced transactions unsubstantiated.

\subsection{Putting It All Toghether}\label{sec:putting-together}

\begin{figure}
    \centering
    \includegraphics[width=\linewidth]{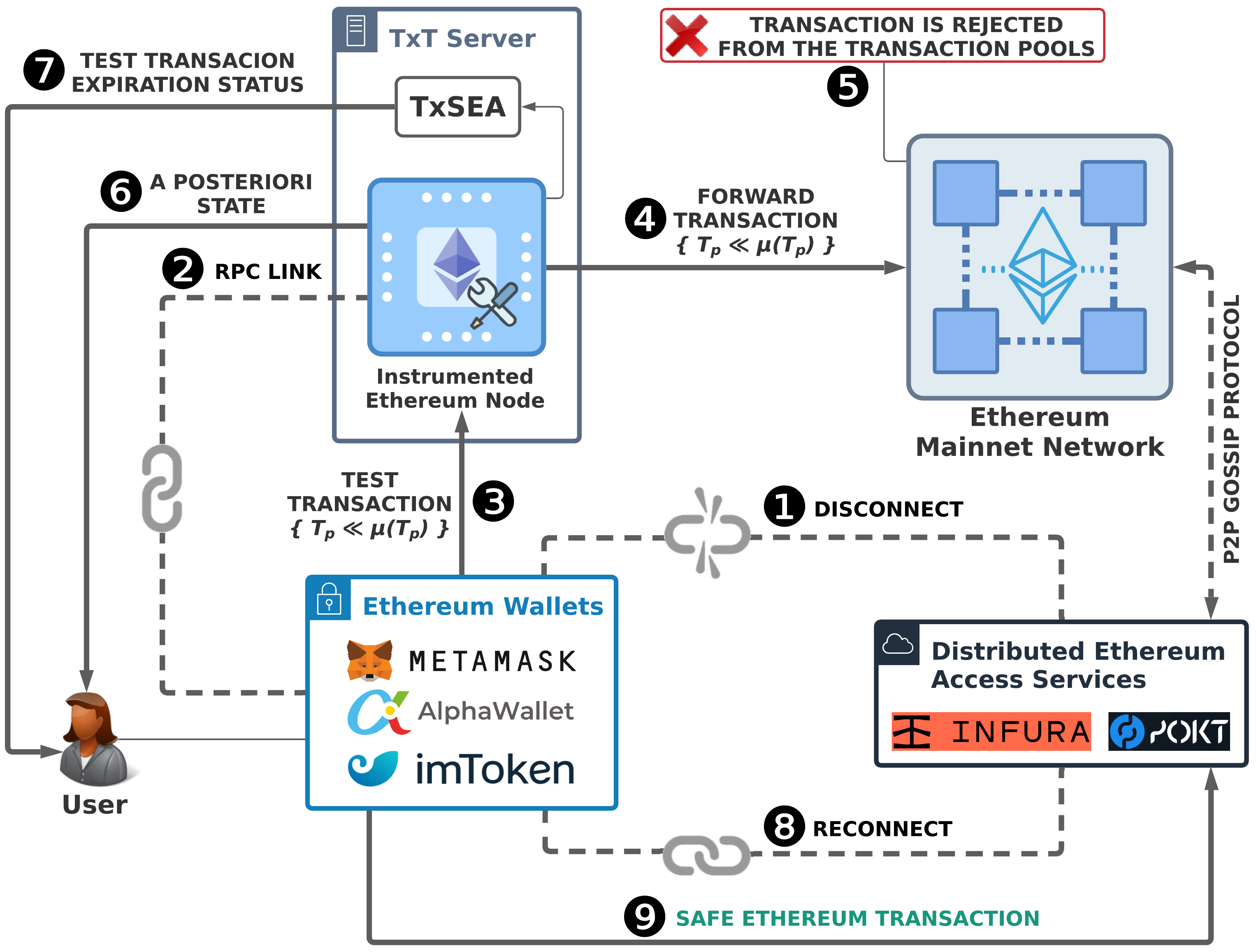}
    \caption{The workflow of TxT testing.}
    \label{fig:txt-overview}
\end{figure}

Fig.~\ref{fig:txt-overview} shows a successful testing of a single transaction. Without the loss of generality, the same workflow can be applied to a series of two or more transactions. We assume that the user has an Ethereum wallet with an account and some positive Ether balance. By specifying the minimal positive gas price of 1 wei/gas, it would require the user to have only $1.35 \cdot 10^{-7}$ USD worth of Ether (as of November 2021) for a worst-case transaction consuming the entire block gas limit.
We also assume that the user submits a transaction either directly to a smart contract, or uses a dApp as a front-end for a smart contract (with the wallet connected to this dApp). Next, we describe each of the nine steps of a successful transaction security testing using TxT.

\ding{182}\xspace\emph{Unplugging from Distributed Node:}
Virtually all Ethereum wallets are connected to Mainnet using a distributed Ethereum Access Service, such as Infura~\cite{infura} or POKT~\cite{pokt}. In order to test transactions with TxT, the user should connect to the TxT server. 

\ding{183}\xspace\emph{Connecting to TxT Node:}
Popular advanced Ethereum wallets, such as MetaMask, allow to connect to a custom Ethereum network by providing its address and port. In most wallets, this is a one-time setup, after which the user can use a drop-down menu to switch between TxT service and Mainnet.

\ding{184}\xspace\emph{Sending Test Transaction:}
Once the user switches to the the TxT network, which is essentially the Mainnet network accessed through the TxT Ethereum node, the user submits a transaction as if it was a usual transaction. This prompts the wallet to show the confirmation dialog, asking the user to select or manually enter the fee parameters. The user specifies a very low gas price (e.g., 1 wei)\footnote{Some wallets prohibit tiny gas prices for Mainnet transactions. However, they do not impose gas price limits on TxT because it is a custom network.}. Once the transaction is submitted, TxT immediately begins processing it.

\ding{185}\xspace\emph{Transaction Forwarding:}
Next, the instrumented node forwards the transaction to the Ethereum Mainnet network using the gossip P2P protocol. Since we aim at preventing the execution of the test transaction by the Mainnet network at large, we expect the transaction to be rejected by all other nodes except the instrumented TxT node due to a very low gas price (i.e., $T_p \ll \mu(T_p)$) specified by the user in the wallet.

\ding{186}\xspace\emph{Rejection by Mainnet at Large:}
Ethereum nodes place transactions into \emph{transaction pools}, in which transactions are awaiting execution. Our experiments confirm that severely underpriced transactions are rejected by most Ethereum nodes early on, without reaching the transaction pools.

\ding{187}\xspace\emph{A Posteriori State:}
Since the user's wallet is connected to the TxT network, the state of TxT node becomes the ground truth for the wallet or a dApp connected to that wallet. Therefore, the test transaction rejected by the Mainnet network outside of the TxT node will be seen as executed by the wallet. We call this situation the \emph{a posteriori state}, i.e., the state of the blockchain caused by the execution of the test transaction.

\ding{188}\xspace\emph{Test Transaction Status:}
TxT provides the status information for each transaction (delivered via a web page, API, or other methods). After submitting the transaction, the user will observe one of the following four test transaction statuses: \textbf{S1:} Transaction is unconditionally testable (\sd) and valid (unexpired); \textbf{S2:} Transaction is \sd, but it is expired; \textbf{S3:} Transaction is \snd, but TxT found a potential vulnerability; and \textbf{S4:} Transaction is untestable (\snd and no vulnerabilities found).

If the transaction is successfully executed and unexpired (\textbf{S1}), the user may submit the test transaction to Mainnet, \emph{if the a posteriori state is satisfactory}. If the transaction is successfully executed but expired (\textbf{S2}), then the user should repeat the test.
If the transaction is \snd with a potential vulnerability warning (\textbf{S3}), the user cannot rely on TxT for testing the transaction, but TxT provides a warning facilitating the assessment of risks via traditional methods. Finally, if TxT determines that the transaction is untestable (\textbf{S4}) that the transaction cannot be evaluated by TxT.

\ding{189}\xspace\emph{Reconnecting to Distributed Node:}
If the transaction is testable, unexpired, and a posteriori state matches the user expectation, then this transaction can be safely submitted to Mainnet for final execution. In this case, the user switches the wallet back to the Mainnet network node for submitting the transaction as usual.

\ding{190}\xspace\emph{Submitting Mainnet Transaction:}
An unexpired \sd transaction is guaranteed to have the same outcome as the test transaction. Even if transaction expires right at the moment it is submitted, the user might initiate an emergency cancellation of the transaction before it is mined following the Ethereum transaction replacement procedure supported by most crypto wallets~\cite{canceltx}.

The above procedure corroborates that a TxT user does not require to employ advanced technical skills (e.g., understanding the contract code) or
meticulously investigate the safety of a planned transaction (or transaction sequence). Moreover, the user assesses the outcome of the test transaction(s) using the familiar interfaces, such as crypto wallet and/or dApp.
\section{Implementation and Evaluation}\label{sec:evaluation}
In this section, we evaluate our implementation of TxT to confirm the feasibility of its real-world deployment.

\begin{table*}
\setlength{\tabcolsep}{3.8pt}
    \centering
    \caption{\rev{Summary of vulnerability coverage by state-of-the-art defense tools.}}\label{table:coverage}
    
\begin{tabular}{c||p{1.0mm}p{1.0mm}p{1.0mm}p{1.0mm}p{1.0mm}p{1.0mm}p{1.0mm}p{1.0mm}p{1.0mm}p{1.0mm}p{1.0mm}p{1.0mm}p{1.0mm}p{1.0mm}p{1.0mm}p{1.0mm}p{1.0mm}p{1.0mm}p{1.0mm}p{1.0mm}p{1.0mm}p{1.0mm}p{1.0mm}p{1.0mm}p{1.0mm}p{1.0mm}p{1.0mm}p{1.0mm}p{1.0mm}p{1.0mm}p{1.0mm}p{1.0mm}p{1.0mm}p{1.0mm}p{1.0mm}p{1.0mm}p{1.0mm}}
        
        \arrayrulecolor{red}\toprule
        
        \multicolumn{1}{c}{} &
        \multicolumn{37}{c}{\textit{Vulnerability (SWC Registry number)}$\mathrm{^\dagger}$} \\
        \cmidrule(lr){2-38}

        \textbf{Defense} &
        \scriptsize \parbox[t]{1mm}{\multirow{2}{*}{\rotatebox[origin=c]{90}{ 100 }}} &
        \scriptsize \parbox[t]{1mm}{\multirow{2}{*}{\rotatebox[origin=c]{90}{ 101 }}} &
        \scriptsize \parbox[t]{1mm}{\multirow{2}{*}{\rotatebox[origin=c]{90}{ 102 }}} &
        \scriptsize \parbox[t]{1mm}{\multirow{2}{*}{\rotatebox[origin=c]{90}{ 103 }}} &
        \scriptsize \parbox[t]{1mm}{\multirow{2}{*}{\rotatebox[origin=c]{90}{ 104 }}} &
        \scriptsize \parbox[t]{1mm}{\multirow{2}{*}{\rotatebox[origin=c]{90}{ 105 }}} &
        \scriptsize \parbox[t]{1mm}{\multirow{2}{*}{\rotatebox[origin=c]{90}{ 106 }}} &
        \scriptsize \parbox[t]{1mm}{\multirow{2}{*}{\rotatebox[origin=c]{90}{ 107 }}} &
        \scriptsize \parbox[t]{1mm}{\multirow{2}{*}{\rotatebox[origin=c]{90}{ 108 }}} &
        \scriptsize \parbox[t]{1mm}{\multirow{2}{*}{\rotatebox[origin=c]{90}{ 109 }}} &
        \scriptsize \parbox[t]{1mm}{\multirow{2}{*}{\rotatebox[origin=c]{90}{ 110 }}} &
        \scriptsize \parbox[t]{1mm}{\multirow{2}{*}{\rotatebox[origin=c]{90}{ 111 }}} &
        \scriptsize \parbox[t]{1mm}{\multirow{2}{*}{\rotatebox[origin=c]{90}{ 112 }}} &
        \scriptsize \parbox[t]{1mm}{\multirow{2}{*}{\rotatebox[origin=c]{90}{ 113 }}} &
        \scriptsize \parbox[t]{1mm}{\multirow{2}{*}{\rotatebox[origin=c]{90}{ 114 }}} &
        \scriptsize \parbox[t]{1mm}{\multirow{2}{*}{\rotatebox[origin=c]{90}{ 115 }}} &
        \scriptsize \parbox[t]{1mm}{\multirow{2}{*}{\rotatebox[origin=c]{90}{ 116 }}} &
        \scriptsize \parbox[t]{1mm}{\multirow{2}{*}{\rotatebox[origin=c]{90}{ 117 }}} &
        \scriptsize \parbox[t]{1mm}{\multirow{2}{*}{\rotatebox[origin=c]{90}{ 118 }}} &
        \scriptsize \parbox[t]{1mm}{\multirow{2}{*}{\rotatebox[origin=c]{90}{ 119 }}} &
        \scriptsize \parbox[t]{1mm}{\multirow{2}{*}{\rotatebox[origin=c]{90}{ 120 }}} &
        \scriptsize \parbox[t]{1mm}{\multirow{2}{*}{\rotatebox[origin=c]{90}{ 121 }}} &
        \scriptsize \parbox[t]{1mm}{\multirow{2}{*}{\rotatebox[origin=c]{90}{ 122 }}} &
        \scriptsize \parbox[t]{1mm}{\multirow{2}{*}{\rotatebox[origin=c]{90}{ 123 }}} &
        \scriptsize \parbox[t]{1mm}{\multirow{2}{*}{\rotatebox[origin=c]{90}{ 124 }}} &
        \scriptsize \parbox[t]{1mm}{\multirow{2}{*}{\rotatebox[origin=c]{90}{ 125 }}} &
        \scriptsize \parbox[t]{1mm}{\multirow{2}{*}{\rotatebox[origin=c]{90}{ 126 }}} &
        \scriptsize \parbox[t]{1mm}{\multirow{2}{*}{\rotatebox[origin=c]{90}{ 127 }}} &
        \scriptsize \parbox[t]{1mm}{\multirow{2}{*}{\rotatebox[origin=c]{90}{ 128 }}} &
        \scriptsize \parbox[t]{1mm}{\multirow{2}{*}{\rotatebox[origin=c]{90}{ 129 }}} &
        \scriptsize \parbox[t]{1mm}{\multirow{2}{*}{\rotatebox[origin=c]{90}{ 130 }}} &
        \scriptsize \parbox[t]{1mm}{\multirow{2}{*}{\rotatebox[origin=c]{90}{ 131 }}} &
        \scriptsize \parbox[t]{1mm}{\multirow{2}{*}{\rotatebox[origin=c]{90}{ 132 }}} &
        \scriptsize \parbox[t]{1mm}{\multirow{2}{*}{\rotatebox[origin=c]{90}{ 133 }}} &
        \scriptsize \parbox[t]{1mm}{\multirow{2}{*}{\rotatebox[origin=c]{90}{ 134 }}} &
        \scriptsize \parbox[t]{1mm}{\multirow{2}{*}{\rotatebox[origin=c]{90}{ 135 }}} &
        \scriptsize \parbox[t]{1mm}{\multirow{2}{*}{\rotatebox[origin=c]{90}{ 136 }}} \\
        \textbf{Tool} & & & & & & & & & & & & & & & & & & & & & & & & & & & & & & & & & & & & & \\
        \arrayrulecolor{red}\midrule
        

        \gc Oyente~\cite{luu2016making} & \gc \circI & \gc \circI & \gc \circI & \gc \circI & \gc \circI & \gc \circI &
        \gc \circI & \gc \circIII & \gc \circI & \gc \circI & \gc \circI & \gc \circI & \gc \circI & \gc \circIII & \gc \circIII & \gc \circI & \gc \circIII & \gc \circI & \gc \circI & \gc \circI & \gc \circI & \gc \circI & \gc \circI & \gc \circI & \gc \circI & \gc \circI & \gc \circI & \gc \circI & \gc \circI & \gc \circI & \gc \circI & \gc \circI & \gc \circI & \gc \circI & \gc \circI & \gc \circI & \gc \circI ~ \\
        
        Securify~\cite{tsankov2018securify} & \circI & \circI & \circI & \circI & \circIII & \circII & \circIII & \circIII & \circI & \circI & \circI & \circI & \circI & \circI & \circIII & \circI & \circI & \circI & \circI & \circI & \circI & \circI & \circI & \circIII & \circIII & \circI & \circI & \circI & \circI & \circI & \circI & \circI & \circI & \circI & \circI & \circI & \circI ~ \\
        
        \gc Mythril~\cite{mueller2018smashing} & \gc \circI & \gc \circIII & \gc \circI & \gc \circI & \gc \circIII & \gc \circI & 
        \gc \circIII & \gc \circIII & \gc \circI & \gc \circI & \gc \circIII & \gc \circI & \gc \circIII & \gc \circI & \gc \circIII & \gc \circI & \gc \circIII & \gc \circI & \gc \circII & \gc \circI & \gc \circI & \gc \circI & \gc \circI & \gc \circI & \gc \circI & \gc \circI & \gc \circI & \gc \circI & \gc \circI & \gc \circI & \gc \circI & \gc \circI & \gc \circI & \gc \circI & \gc \circI & \gc \circI & \gc \circI ~
        \\
        
        Sereum~\cite{rodler2018sereum} & \circI & \circI & \circI & \circI & \circI & \circI &
        \circI & \circIII & \circI & \circI & \circI & \circI & \circI & \circI & \circI & \circI & \circI & \circI & \circI & \circI & \circI & \circI & \circI & \circI & \circI & \circI & \circI & \circI & \circI & \circI & \circI & \circI & \circI & \circI & \circI & \circI & \circI ~
        \\        
        
        \gc Vandal~\cite{brent2018vandal} & \gc \circI & \gc \circI & \gc \circI & \gc \circI & \gc \circI & \gc \circIII &
        \gc \circIII & \gc \circIII & \gc \circI & \gc \circI & \gc \circI & \gc \circI & \gc \circI & \gc \circI & \gc \circI & \gc \circIII & \gc \circI & \gc \circI & \gc \circIII & \gc \circI & \gc \circI & \gc \circI & \gc \circI & \gc \circI & \gc \circI & \gc \circI & \gc \circI & \gc \circI & \gc \circI & \gc \circI & \gc \circI & \gc \circI & \gc \circI & \gc \circI & \gc \circII & \gc \circI & \gc \circI ~
        \\
        
        sGuard~\cite{nguyen2021sguard} & \circI & \circIII & \circI & \circI & \circI & \circI &
        \circI & \circIII & \circI & \circI & \circI & \circI & \circI & \circI & \circI & \circIII & \circI & \circI & \circI & \circI & \circI & \circI & \circI & \circI & \circI & \circI & \circI & \circI & \circI & \circI & \circI & \circI & \circI & \circI & \circI & \circI & \circI ~
        \\        
        
        \gc ZEUS~\cite{kalra2018zeus} & \gc \circI & \gc \circIII & \gc \circI & \gc \circI & \gc \circI & \gc \circI &
        \gc \circI & \gc \circIII & \gc \circI & \gc \circI & \gc \circI & \gc \circI & \gc \circI & \gc \circI & \gc \circIII & \gc \circIII & \gc \circIII & \gc \circI & \gc \circI & \gc \circI & \gc \circI & \gc \circI & \gc \circI & \gc \circI & \gc \circI & \gc \circI & \gc \circI & \gc \circI & \gc \circI & \gc \circI & \gc \circI & \gc \circI & \gc \circI & \gc \circI & \gc \circIII & \gc \circI & \gc \circI ~
        \\
        
        ConFuzzius~\cite{ferreira2021confuzzius} &  \circI &  \circIII &  \circI &  \circI &  \circIII &  \circI & 
         \circIII &  \circIII &  \circI &  \circI &  \circIII &  \circI &  \circIII &  \circI &  \circIII &  \circI &  \circIII &  \circI &  \circII &  \circI &  \circI &  \circI &  \circI &  \circI &  \circI &  \circI &  \circI &  \circI &  \circI &  \circI &  \circI &  \circI &  \circI &  \circI &  \circI &  \circI &  \circI ~
        \\        
        
        \gc VeriSmart~\cite{so2020verismart} & \gc \circI & \gc \circIII & \gc \circI & \gc \circI & \gc \circI & \gc \circI &
        \gc \circI & \gc \circI & \gc \circI & \gc \circI & \gc \circI & \gc \circI & \gc \circI & \gc \circI & \gc \circI & \gc \circI & \gc \circI & \gc \circI & \gc \circI & \gc \circI & \gc \circI & \gc \circI & \gc \circI & \gc \circI & \gc \circI & \gc \circI & \gc \circI & \gc \circI & \gc \circI & \gc \circI & \gc \circI & \gc \circI & \gc \circI & \gc \circI & \gc \circI & \gc \circI & \gc \circI ~
        \\
        
        SmarTest~\cite{so2021smartest} & \circI & \circIII & \circI & \circI & \circI & \circIII &
        \circIII & \circI & \circI & \circI & \circIII & \circI & \circI & \circI & \circI & \circI & \circI & \circI & \circI & \circI & \circI & \circI & \circI & \circII & \circI & \circI & \circI & \circI & \circI & \circI & \circI & \circI & \circI & \circI & \circI & \circI & \circI ~
        \\

        \gc \rev{Osiris~\cite{torres2018osiris}} & \gc \circI & \gc \circIII & \gc \circI & \gc \circI & \gc \circI & \gc \circI &
        \gc \circI & \gc \circI & \gc \circI & \gc \circI & \gc \circI & \gc \circI & \gc \circI & \gc \circI & \gc \circI & \gc \circI & \gc \circI & \gc \circI & \gc \circI & \gc \circI & \gc \circI & \gc \circI & \gc \circI & \gc \circI & \gc \circI & \gc \circI & \gc \circI & \gc \circI & \gc \circI & \gc \circI & \gc \circI & \gc \circI & \gc \circI & \gc \circI & \gc \circI & \gc \circI & \gc \circI ~ 
        \\
        
        \rev{ECFChecker~\cite{grossman2017online}} & \circI & \circI & \circI & \circI & \circI & \circI &
        \circI & \circIII & \circI & \circI & \circI & \circI & \circI & \circI & \circI & \circI & \circI & \circI & \circI & \circI & \circI & \circI & \circI & \circI & \circI & \circI & \circI & \circI & \circI & \circI & \circI & \circI & \circI & \circI & \circI & \circI & \circI ~
        \\

        \gc \rev{Maian~\cite{nikolic2018finding}} & \gc \circI & \gc \circI & \gc \circI & \gc \circI & \gc \circI & \gc \circIII &
        \gc \circIII & \gc \circI & \gc \circI & \gc \circI & \gc \circI & \gc \circI & \gc \circI & \gc \circI & \gc \circI & \gc \circI & \gc \circI & \gc \circI & \gc \circI & \gc \circI & \gc \circI & \gc \circI & \gc \circI & \gc \circI & \gc \circI & \gc \circI & \gc \circI & \gc \circI & \gc \circI & \gc \circI & \gc \circI & \gc \circI & \gc \circI & \gc \circI & \gc \circI & \gc \circI & \gc \circI ~
        \\
        
        \textbf{TxT (this work)} & \circIII & \circIII & \circIII & \circIII & \circIII & \circI &
        \circIII & \circIII & \circIII & \circIII & \circIII & \circIII & \circIII & \circIII & \circI & \circIII & {\tiny \faWarning} & \circIII & \circIII & \circIII & {\tiny \faWarning} &  \circIII & \circIII & \circIII & \circIII & \circIII & \circIII & \circIII & \circIII & \circIII & \circIII & \circIII & \circIII & \circIII & \circI & \circIII & \circI
        \\ 
                        
        \arrayrulecolor{red}\bottomrule
        
        \multicolumn{38}{c}{\scriptsize
        \CIRCLE~--- full support; \LEFTcircle~--- partial support; \Circle~--- no support; {\tiny \faWarning}~--- explicit detection of vulnerability.
        } \\
        
        \arrayrulecolor{red}\hline
        
        \multicolumn{38}{l}{$\mathrm{^\dagger}$ https://swcregistry.io/.}
        
    \end{tabular}
\end{table*}

\subsection{Implementation and Deployment}

We implement TxT by instrumenting Go Ethereum 1.10.10 and adding additional data-processing modules using Node.js 12.22.5 with Web3.js 1.2.6 and Python 3.9.7. In order to prevent accidental disruption of the normal Ethereum execution, our instrumentation of Go Ethereum includes only minimal necessary modifications, i.e., gathering and saving chain data, and overriding the gas price bottom limitations \emph{only for specified accounts} representing the customers of a TxT server. The gathered data is then processed independently of the node by external Node.js and Python modules.

We deploy TxT on Dell PowerEdge T640 server with 2  Intel Xeon Gold 5218 CPU, 250 GB RAM, and SATA SSD (6 Gbps throughput), connected to 1 Gbps wired Internet link. The instrumented TxT node uses the full synchronization mode with one CPU mining thread (for enabling opcode execution), and 8,192 MB of cache. In current implementation, we use SSH and text interface for test transaction status retrieval. 

\subsection{Vulnerability Coverage by TxT}\label{sec:eval-coverage}

We implement 37 cases reproducing all the cataloged smart contract vulnerabilities in the SWC Registry~\cite{swcregistry}.
After that, we test all the transaction sequences reproducing these vulnerabilities on a TxT deployment to assess which vulnerabilities are detectable by TxT. One important aspect of this assessment is that we judge the ability of TxT to reveal a vulnerability not only based on our sample implementation, but also based on the ability to address \emph{all possible vulnerable implementations}. \rev{We compare TxT with 13 state-of-the-art defenses based on their self-reported coverage disclosure.}

The result, shown in Table~\ref{table:coverage}, demonstrates that TxT significantly outperforms all the state-of-the art tools in terms of the number of vulnerabilities it is able to defend against. Specifically, all the state-of-the-art tools combined only detect and/or prevent 15 out of 37 vulnerabilities (40.5\% coverage), while TxT \emph{deterministically} prevents 31 out of 37 (83.8\% coverage). Furthermore, if we add the warnings of potential insecurity to our assessment, the vulnerability coverage by TxT reaches 89.2\%.

Some vulnerabilities, such as SWC-105, SWC-115 and SWC-134, are semantic-dependent, i.e., they rely upon understanding of the intent of the developer and/or user, and therefore they are only supported by the heuristic tools. For example, a pattern corresponding to SWC-105 (``Unprotected Ether Withdrawal'') is perceived as a dangerous omission in most contracts, but the same behavior could be correct if the contract is designed to be an Ether faucet.
Moreover, SWC-136 (``Unencrypted Private Data On-Chain'') still remains unsupported by all existing tools, including TxT. Addressing this vulnerability would requires the identification of a leaked secret, which is insurmountable.  

\subsection{Transaction Expiration Rate}\label{sec:eval-expiration}
TxT tests are prone to expiration due to the constantly changing state of blockchain. In this evaluation, we gather over 1.3 billion transactions (from the Genesis block until November 5, 2021) submitted to over 131 million Ethereum accounts (smart contracts and EOAs) to find the percentage of accounts resilient to transaction expiration. To assess the transaction expiration resiliency, we pick three time thresholds (1 minute, 10 minutes, and 1 hour), and we group all accounts into three categories: 1) the ones that have never experienced transaction expiration within the set threshold; 2) the ones which transactions on average (mean) do not expire before the threshold; and 3) the ones with 90\% or more transactions not expiring before the set threshold, as shown in Table~\ref{tab:retention}.

\begin{table}
\setlength{\tabcolsep}{4pt}
    \centering
    
    \caption{Number ($\times 10^6$) and percentage of accounts exhibiting state retention within set time threshold.
    }
    \label{tab:retention}
    \begin{tabular}{c||c:c:c}
        
        \arrayrulecolor{red}\toprule
        
        \multicolumn{1}{c}{\multirow{2}{*}{\textbf{Counting}}} &
        \multicolumn{3}{c}{\textit{State retention threshold ($\theta_{exp}$)}} \\
        \cmidrule(lr){2-4}

        \textbf{condition} &
        60 sec. & 600 sec. & 3600 sec. \\
        \arrayrulecolor{red}\midrule
        

        All txns testable & 122.38 & 115.11 &  109.68 \\
        ($min(\Delta t) > \theta_{exp}$)& (93.19\%) & (87.65\%) & (83.52\%) \\
        \hline

        Avg. txns testable & 124.80 & 121.50 &  119.08 \\
        ($\mu(\Delta t) > \theta_{exp}$) & (95.04\%) & (92.52\%) & (90.67\%) \\
        \hline

        90\% testable & 124.86 & 121.59 &  119.19 \\
        ($P_{90\%}(\Delta t) > \theta_{exp}$) & (95.08\%) & (92.59\%) & (90.76\%) \\

        \arrayrulecolor{red}\bottomrule
        
        
        
        
    \end{tabular}
\end{table}

\begin{figure}
    \centering
    \includegraphics[width=2.4in]{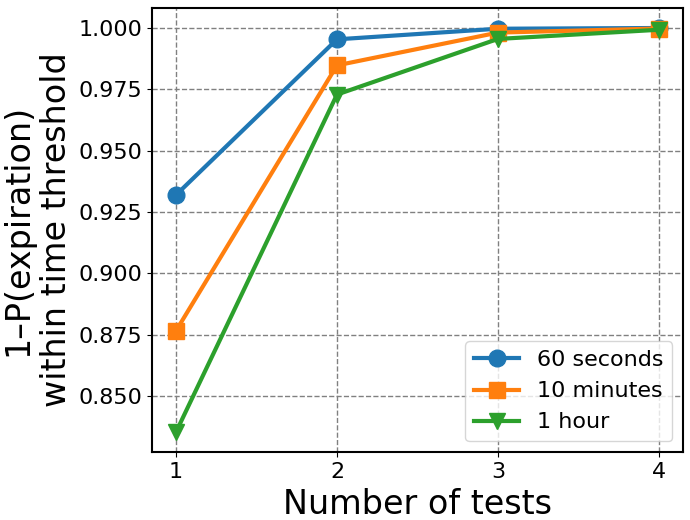}
    \caption{Probability of avoiding expiration via repeated testing.}
    \label{fig:exp}
\end{figure}

The experimental result demonstrates that statistically the vast majority of test transactions will not expire within reasonable time, sufficient for submitting the final transaction to Mainnet. However, if the test expires earlier than the final Mainnet transaction is submitted, the user has a choice to repeat the test, and the probability of success after the multiple tests will be $P_{succ} = 1 - (1 - P_{single})^{k}$, where $P_{single}$ is the probability of success for a single test within the given time threshold, and $k$ is the number of attempts. Thus, even if transaction expires before the user submits the final one, a repeated test will  address the problem, as shown in Fig.~\ref{fig:exp}.

\subsection{$\sigma$-nondeterministic Transactions}\label{sec:eval-tbd}

In this work, we propose a paradigm allowing to deterministically predict the result of a transaction at the expense of rejecting a small portion of transactions that we call \snd. TxT is unable to guarantee the outcome of a \snd transaction, however, we are able to partition these transactions into potentially unsafe (prone to SWC-120 and SWC-116 vulnerabilities), and untestable (not necessarily vulnerable, but the result is unpredictable).

\begin{figure}
    \centering
    \includegraphics[width=0.9\linewidth]{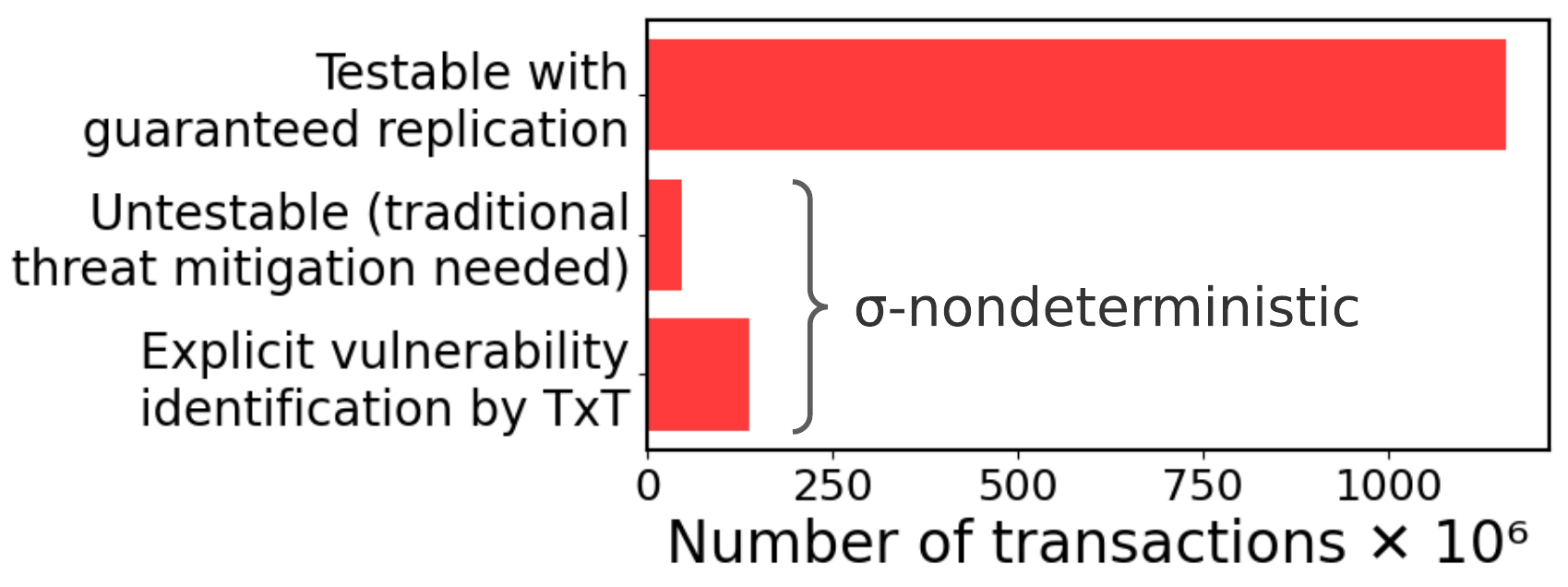}
    \caption{Occurrence of \snd opcodes in the execution stack of 1.3 billion 
    Ethereum transactions.}
    \label{fig:opcodes2}
\end{figure}

Fig.~\ref{fig:opcodes2} shows the result of processing over 1.3 billion Ethereum transactions with opcode analysis of their execution stacks. The result shows the counts of opcode presence events (i.e., several identical opcodes within one call stack count as one event), divided into three groups: untestable (no vulnerability markers), SWC-120 markers, and SWC-116 markers. The latter two groups produce respective warnings regarding possible vulnerabilities, while the untestable transactions are to be rejected by TxT.

The evaluation shows that approximately 86.27\% of all transactions are \sd and 13.73\% are \snd. Out of almost 185 million \snd transactions, only 25.5\% are purely untestable, which means that TxT completely rejects about 3.5\% of transactions, and gives at least partial results for 96.5\% of transactions. We believe that through a deep opcode and EVM stack analysis it is possible to further reduce the rate of \snd and untestable transactions. 

\subsection{Underpriced Transactions in the Wild}\label{sec:eval-accidental}

\begin{figure}
    \centering
    \includegraphics[width=2.2in]{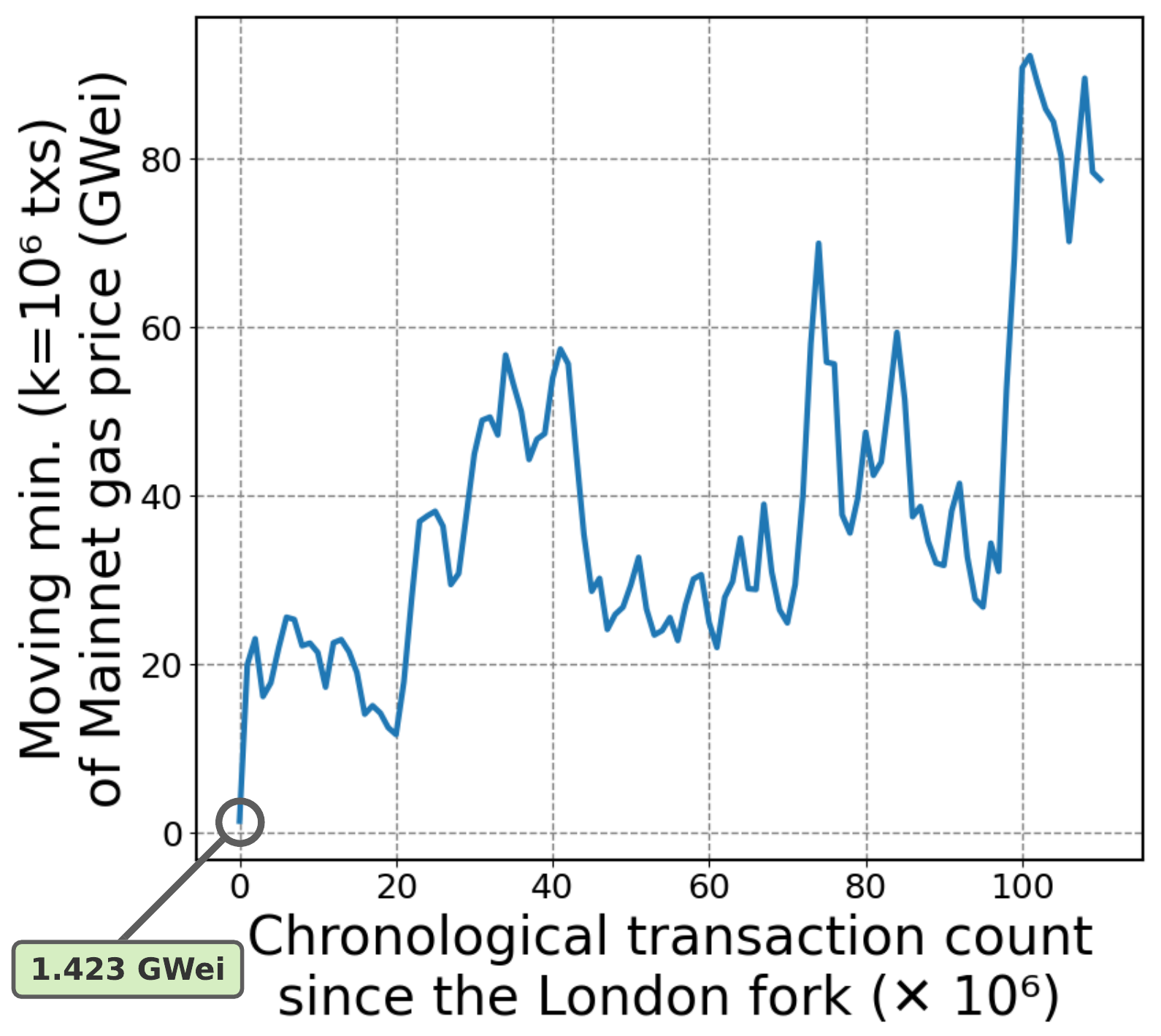}
    \caption{Minimal accepted gas price of 111,226,625 post-London transactions on Mainnet. Without the loss of correctness, we apply the moving minimum function to the data.}
    \label{fig:gasprice2}
\end{figure}

The transaction underpricing approach, utilized by TxT, raises a concern: if a block does not have enough properly priced competing transactions, the underpriced test transaction might be included in this block~\cite{wood2014ethereum}. Our evaluation shows that Ethereum Mainnet has 2,506,498 zero-priced transactions (as of November 2021)\footnote{For example, \texttt{0xc3fa8399ef7922aef0ec7278f7b4b5e28e7191e} \texttt{ba3027ca1143af2cf17acae86}}. These transactions have been a known nuisance in the Ethereum community~\cite{zero-gas}.
Although the rate of zero-priced transactions on Ethereum is only 0.186\%, the very fact of their presence poses a
threat to the feasibility of TxT. Fortunately, the EIP-1559~\cite{buterin2019eip} proposal, which has been enforced at the London hard fork, solves the problem. Although the protocol adjustment does not explicitly target the zero-priced transactions, it effectively makes these transactions impossible. To verify this, we process over 111 million transactions soon after the London fork to confirm that none of them has a gas price lower than 1,423,420,054 wei (see Fig.~\ref{fig:gasprice2}). Thus, after the London fork, it is no longer possible to accidentally mine an underpriced transaction.

\rev{
\subsection{TxT Delays and Transaction Efficiency}\label{sec:delays}

TxT is implemented as an instrumented Go Ethereum node incurring some additional transaction execution delays. Moreover, as TxT continues processing transactions, the transaction execution delay may increase due to the growing cache size of TxSEA algorithm. In this part of the evaluation, we first measure the added per-opcode delay of TxT instrumentation over a large time period. Then, we make a projection of the added delay of transaction execution by a TxT node.

For our evaluation experiment, we compare the opcode execution delays between instrumented TxT node and a pure Go Ethereum node. The experiment was conducted on the same Dell PowerEdge T640 server as the rest of the experiments. The time-critical core module of TxT is repeatedly invoked in the opcode processing loop of the \texttt{Run} function in \texttt{core/vm/interpreter.go}. We activate TxT for historical Mainnet transactions, and collect timestamps at every iteration of the loop, which gives us the delay of execution of a single opcode. After that, we remove all the TxT code from the node, leaving only the timestamp collection instruction, and execute the same transactions again, this time without TxT. To be able to better visualize the data without the loss of generality, we collect the execution delays of 500 million of executed opcodes, both with and without TxT, and split them into 500 frames, each containing 1 million transactions. Then, for each of the frames, we plot the difference between the average instrumented and non-instrumented delays, which we call the \emph{added delay} (i.e., the difference in opcode processing delay between TxT and baseline approach, see Fig.~\ref{fig:delay1}). The result shows that despite growing TxSEA cache, TxT does not exhibit any noticeable growth in added opcode processing delay. Moreover, the average delay for each frame stays between 2,300 and 3,000 nanoseconds per opcode.

\begin{figure}
    \centering
    \includegraphics[width=0.9\linewidth]{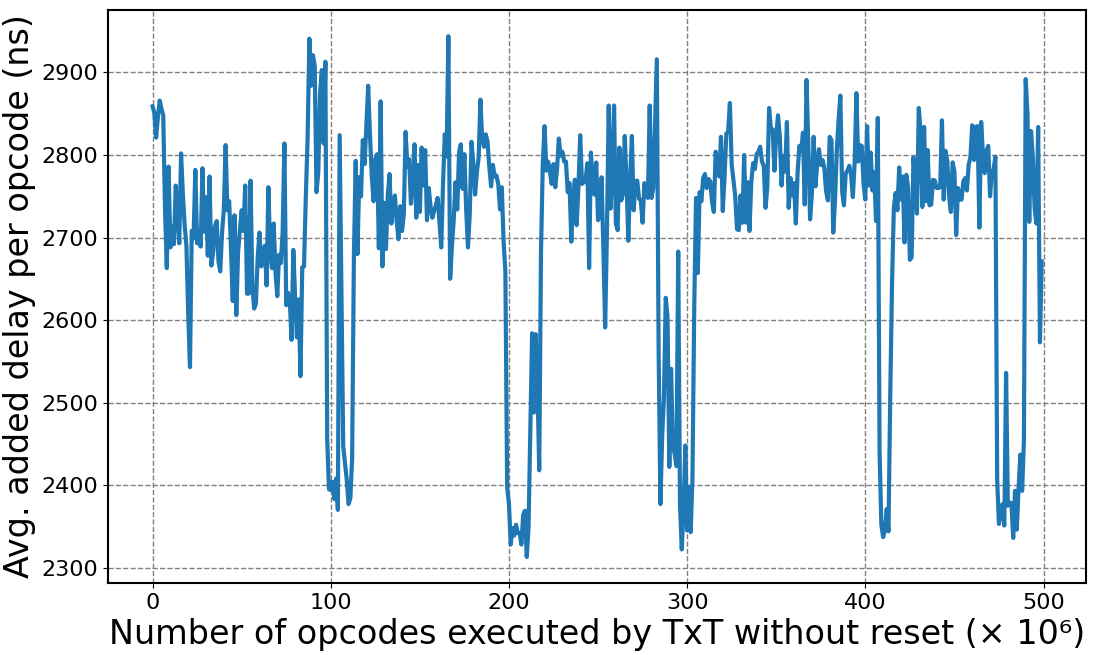}
    \caption{\rev{Additional time (in nanoseconds) that the instrumented TxT node spends on average for executing one opcode compared to the baseline non-instrumented node. Despite growing cache size, the execution delay is not visibly increasing even after 500 million processed opcodes. The measurement is a average of 500 frames, each with 1 million transactions.}}
    \label{fig:delay1}
\end{figure}


In our next experiment, we count the number of opcodes executed by a sample of 100 million transactions in Ethereum Mainnet. Then, we create a distribution of the transaction opcode counts, shown in Fig.~\ref{fig:delay2}. As we can see, the vast majority of transactions execute less than 5,000 opcodes. The results of the evaluation show that most added opcode execution delays are under 3,000 nanoseconds, while the vast majority of transactions execute under 5,000 opcodes. Therefore, the added delay caused by TxT implementation does not exceed $3,000 \cdot 5,000 \cdot 10^{-6}$ = 15 ms. Assuming that the sequence of transactions in a tested workflow does not exceed 10, the TxT delay per test will not be larger than 150 ms, which is negligible. The state-of-the-art smart contract tester Confuzzius~\cite{ferreira2021confuzzius}, which claims enhanced time performance compared to previous testers, requires 500-1,000 seconds of time to achieve 75\% instruction coverage. Compared to Confuzzius, TxT delivers almost instant result because it dynamically tests transactions against the current state of blockchain.

\begin{figure}
    \centering
    \includegraphics[width=2.2in]{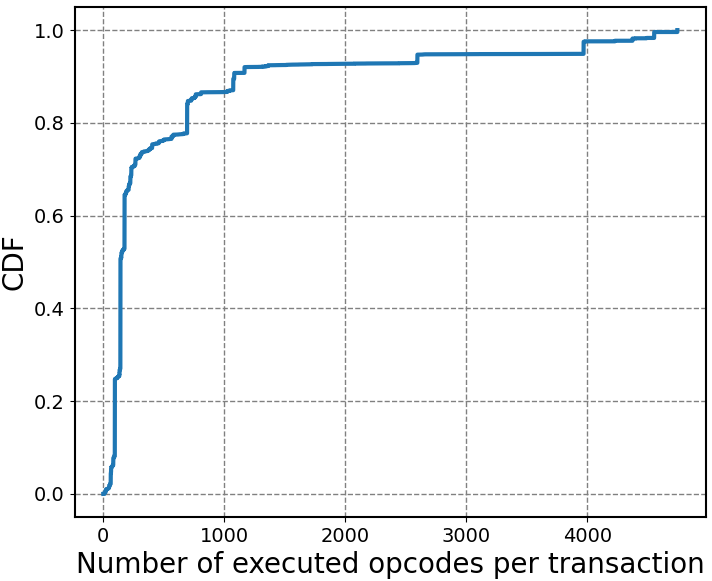}
    \caption{\rev{Number of executed EVM opcodes per transaction, based on the sample of 100 million transactions.}}
    \label{fig:delay2}
\end{figure}



Last, but not least, we evaluated the performance and feasibility of TxT over the Proof-of-Stake consensus that was recently adopted by the Ethereum network. Except for the necessity to update several command-line parameters, we did not notice any performance or other difference between TxT operating on Proof-of-Work consensus versus Proof-of-Stake.
}
\section{Limitations and Discussion}\label{sec:discussion}

This paper is the first work on a deterministic approach of smart contract testing  using the transaction encapsulation. We believe that our work opens up a new era of non-heuristic audit of smart contracts. However, the paradigm shift comes with some remaining challenges and open questions.




\noindent\textbf{Testing Eligibility.}
The current implementation of TxT provides a practical proof of concept of the transaction encapsulation framework. However, the rate of \snd transactions is still  high. Our evaluation shows that the major culprits are the \texttt{NUMBER} and \texttt{BALANCE} opcodes in the call stacks of the tested transactions. However, we observed
that  the conditional statements involving the \texttt{NUMBER} opcode often turn into tautologies or contradictions when $T_{b,i}$ is larger than a certain value. In other words, we have sufficient reason to believe that the rate of \snd transactions can be drastically reduced by designing more fine-tuned procedures for identifying \snd transactions.

\noindent\textbf{State Expiration.}
Blockchain is a dynamic multi-user environment in which executed transactions constantly create interference to one another. This interference is the cause of TxT test expiration. In this work we use a coarse assumption $T_{b,j} > T_{b,i} \wedge T_{t,j} = T_{t,i}$ for determining the transaction expiration. However, we believe we can significantly reduce the rate of expiration by exploring the execution stacks of purportedly interfering transactions and determining which of them \emph{effectively} interfere with the testing transaction.


\noindent\textbf{Custom RPC Support by Cryto Wallets.} 
TxT design assumes that the user wallet, which is the proxy to the Ethereum ecosystem, has the support for adding a custom RPC network. While most popular wallets support this feature, some other wallets (e.g., MyEtherWallet~\cite{mew}) do not. However, in the spirit of decentralization and trust elimination, most Ethereum wallets are open-source, which makes it easy to add a modification or plugin for supporting a custom RPC.

\noindent\textbf{Transactions from Multiple Accounts.}
The current design of TxT assumes that the entire transaction sequence originates from the same account, which is by far the most obvious scenario. However, we also admit that some transaction sequences might require testing involving several accounts, such as in the case of multi-signature distributed token wallets. Although the multi-account support is deliberately removed from the current design to avoid unnecessary complication, our analysis indicates that implementing this functionality is tantamount to improving some bookkeeping routines in the current design.

\noindent\textbf{Deployment Scalability.}
The current implementation of TxT does not allow to run multiple testing sessions on one instrumented node, which means that the TxT security provider must maintain a sufficient number of separate instrumented Ethereum nodes to accommodate all simultaneous testing requests. On the one hand, the computation cost is not a big concern because TxT instrumented nodes do not require competitive mining. On the other hand, each node must be fully synchronized, with approximately 500GB of per-node storage requirement.
Yet, we believe it is possible to orchestrate testing to allow execution of non-conflicting transactions  on the same node. We leave this functionality for future research.

\section{Related Work}\label{sec:relatedwork}
In this section, we show how TxT compares to the existing smart contract security defense methods.

\noindent\textbf{Code-based Defense.}
Code-based defense tools use source code, bytecode and/or ABI maps for finding bugs and vulnerabilities in smart contracts. One of the most popular code-based approaches is symbolic execution, represented by Mythril~\cite{mueller2018smashing}, Oyente~\cite{luu2016making}, and Maian~\cite{nikolic2018finding}. SmarTest~\cite{so2021smartest} uses a language-based model for guiding symbolic execution and generating malicious transaction sequences. Static analyzers and formal verifiers, such as Securify~\cite{tsankov2018securify}, EthBMC~\cite{frank2020ethbmc}, VerX~\cite{permenev2020verx}, and Vandal~\cite{brent2018vandal} attempt to extract semantics and other facts from the code for finding violation of safety patterns. Many static analysis tools zero in on specific security issues. For example, ZEUS~\cite{kalra2018zeus}, Osiris~\cite{torres2018osiris}, and VeriSmart~\cite{so2020verismart} focus on arithmetic bugs; ECFChecker~\cite{grossman2017online}, Sereum~\cite{rodler2018sereum}, and SeRIF~\cite{cecchetti12compositional} address reentrancy; TokenScope~\cite{chen2019tokenscope} targets security issues of ERC-20 tokens. The major drawback of code-based defense approaches is the probabilistic nature of the result, which incurs non-negligible 
false positives/negatives. 
In contrast, TxT provides the user with an \emph{actual outcome} of a transaction applied to the current state of blockchain.


\noindent\textbf{Testers.} Smart contract testers allow to generate and execute transactions to unveil vulnerabilities or semantic violations.
Manual testing methods include tools like Waffle~\cite{waffle} and Solidity Coverage~\cite{solidity-coverage}. In order to enhance the ability of the test tools to reveal vulnerabilities, a number of smart contract fuzzing methods have been proposed, including Harvey~\cite{wustholz2020harvey}, Confuzzius~\cite{ferreira2021confuzzius}, ContractFuzzer~\cite{jiang2018contractfuzzer}, and sFuzz~\cite{nguyen2020sfuzz}. These testing methods try to find transaction parameters that would confirm the safety of a smart contract or reveal a vulnerability. However, the search space for the candidate parameters is usually too large to exhaust all the possible values (the path explosion problem); as a result, the testing methods only use some sample sets of parameters or heuristically determined combinations of parameters --- resulting in overlooked vulnerabilities. Instead of predicting future transaction parameters, TxT tests the exact transactions the user is going to submit.

\noindent\textbf{Transaction-based Defense.} Unlike code-base defense tools, which statically scrutinize source code or bytecode of smart contracts, the transaction-based defense tools analyze historical transactions 
stored in the blockchain, or intercept the incoming transactions in real time. TxSpector~\cite{zhang2020txspector} and EthScope~\cite{wu2020ethscope} deliver frameworks for retrospective vulnerability search using Ethereum transactions. SODA~\cite{chen2020soda} and {\AE}gis~\cite{ferreira2020aegis} are tools for online interception of malicious transactions. 
However, none of the existing transaction-based methods 
provide a definitive result to ensure
the transaction safety. In contrast, TxT is a transaction-based dynamic interceptor that  deterministically verifies the safety of  transactions, or refuses to give an answer in case of uncertainty. Qin et al.~\cite{qin2021quantifying} describe a transaction replay scheme similar to TxT. However, the proposed transaction replay is used by the authors to demonstrate a front-running attack, while we use this method for defense. Moreover, in addition to replaying the transaction, TxT also addresses the notorious TOCTOU challenge by assessing the expiration and replicability of the test transaction.

\section{Conclusion}\label{sec:conclusion}

Traditional software often requires user confirmation of critical operations, such as deleting records or submitting web-based applications. Implementing the same mechanism in smart contracts is notoriously hard due to the notorious TOCTOU issue caused by the ever-changing state of the blockchain. In this work, we provided the first solution to address this problem by allowing
a user to preview and confirm transactions. To make it feasible, 
we formally determined the exact set of conditions for transaction replicability and introduced transaction encapsulation, a new framework for deterministic real-time transaction testing, which uncovers 
the outcome of the intended transactions or transaction sequences. 
Transaction encapsulation could effectively capture the unpredictable behaviors associated with known and zero-day vulnerabilities. We developed and implemented the transaction tester TxT. Through extensive experiments, we demonstrated that TxT
prevents the exploitation of more than twice as many vulnerabilities as covered by the existing defense tools combined. In the spirit of open research, we will make TxT and all the evaluation artifacts open source.





\begin{IEEEbiography}[{\includegraphics[width=1in,height=1.25in,clip,keepaspectratio]{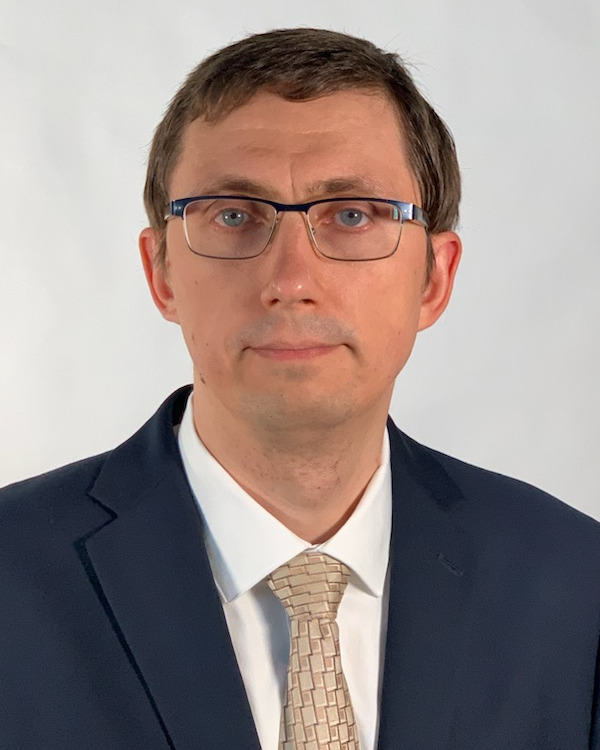}}]{\rev{Nikolay Ivanov}} \rev{is a PhD candidate in the Department of Computer Science and Engineering at Michigan State University. He received his Summa Cum Laude B.Sc degree in Computer Science from Southwest Minnesota State University. He is a Cloud Computing Fellow at Michigan State University. In 2022, he won the Computer Science Outstanding Graduate Student award. His research interests include cybersecurity, smart contract security, applied cryptography, and blockchain-assisted IoT systems.}
\end{IEEEbiography}
\begin{IEEEbiography}[{\includegraphics[width=1in,height=1.25in,clip,keepaspectratio]{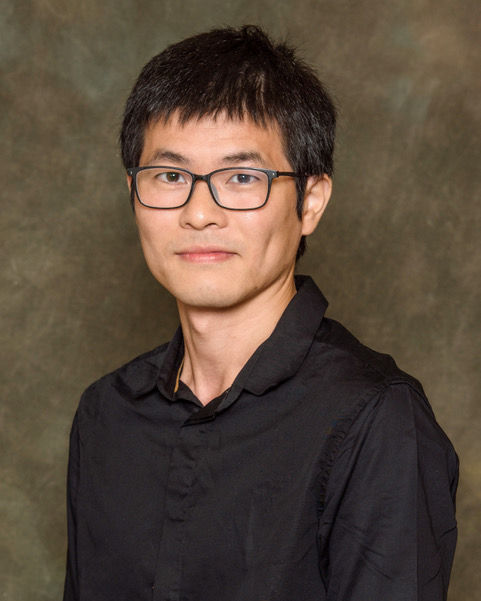}}]{\rev{Qiben Yan}} \rev{is an Assistant Professor in Department of Computer Science and Engineering of Michigan State University. He received his Ph.D. in Computer Science department of Virginia Tech, an M.S. and a B.S. degree in Electronic Engineering from Fudan University in Shanghai, China. He is IEEE Senior Member, and a recipient of NSF CRII award in 2016. His current research interests include wireless communication, wireless network security and privacy, mobile and IoT security, and big data privacy.}
\end{IEEEbiography}
\begin{IEEEbiography}[{\includegraphics[width=1in,height=1.25in,clip,keepaspectratio]{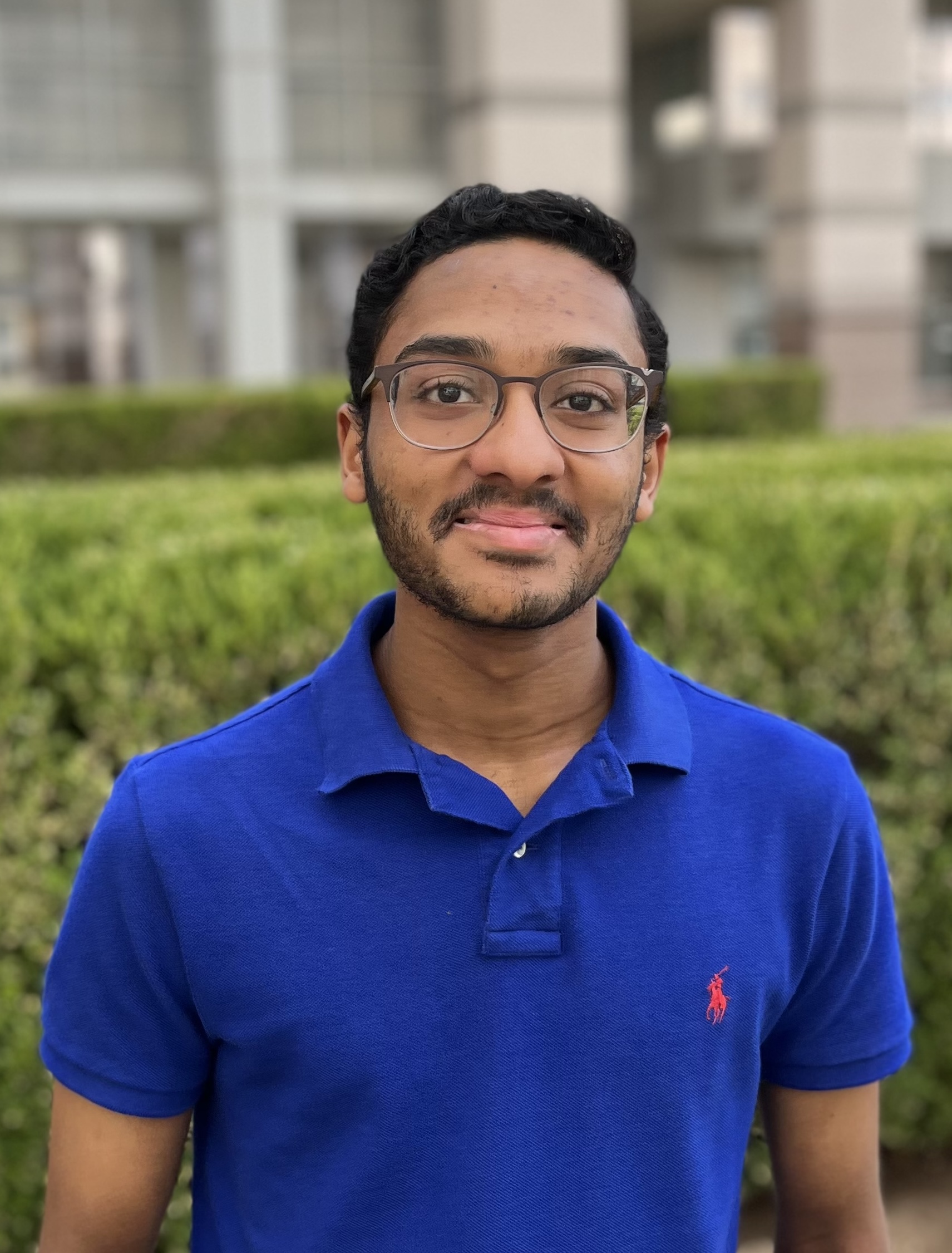}}]{\rev{Anurag Kompalli}} \rev{is an undergraduate student in the College of Computer Science and Engineering at Michigan State University. He is also currently enrolled in the Honors College at Michigan State University and works as an undergraduate research assistant in the Secure Internet of Things Lab at Michigan State. His current research interests are Blockchain and Smart Contract Security, Distributed Systems and Decentralized Finance.}
\end{IEEEbiography}


\ifCLASSOPTIONcaptionsoff
  \newpage
\fi

\end{document}